\documentclass[10pt,twocolumn,aps,prb,superscriptaddress]{revtex4-2}
\usepackage{graphicx,subfigure,physics,dsfont,amsmath,appendix,amssymb,mathrsfs}
\usepackage[utf8]{inputenc}
\usepackage[normalem]{ulem}               
\usepackage{orcidlink}
\usepackage{duckuments}
\usepackage[dvipsnames]{xcolor}           
\usepackage{times,txfonts}
\usepackage{hyperref}
\usepackage{color}
\usepackage{todonotes}
\definecolor{olive}{rgb}{0.01, 0.43, 0.01}
\hypersetup{
    colorlinks=true, linkcolor={red!50!black}, citecolor={blue!80!black},
    urlcolor={purple!80!black}, linktoc=all
}

\newcommand{\dA}{\Delta_a}

\newcommand{\Uab}{U_{ab}}
\newcommand{\Tbeta}{\mathcal{T}_\beta}
\begin{document}

\title{Mobile-impurity dynamics in a non-Hermitian Stark-localised bath}
\author{Muhammad S. Hasan\,\orcidlink{0000-0002-2549-8656}}
\email[]{mshasan.official@gmail.com}
\affiliation{Institute of Atomic and Molecular Sciences, Academia Sinica, Taipei 10617, Taiwan}
\author{Hsiang-Hua Jen\, \orcidlink{0000-0002-1152-8164}}
\email[]{sappyjen@gmail.com}
\affiliation{Institute of Atomic and Molecular Sciences, Academia Sinica, Taipei 10617, Taiwan}
\affiliation{Department of Physics, National Central University, Taoyuan City, 320317, Taiwan}
\affiliation{Physics Division, National Center for Theoretical Sciences, Taipei 10617, Taiwan}
\author{Jhih-Shih You}
\email[]{jhihshihyou@ntnu.edu.tw}
\affiliation{Department of Physics, National Taiwan Normal University, Taipei 11677, Taiwan}

\begin{abstract}
%A linear potential can localise an interacting quantum system without a quenched disorder, while asymmetric hopping produces the well-known non-Hermitian skin effect. 
We investigate the dynamics of a clean mobile impurity coupled to a bath subjected to a linear Stark potential as well as asymmetric hopping in one dimension. The impurity mediates resonant tunnelling in the bath whenever the interaction strength matches the Stark potential and leads to loss in the memory of the initial state of the bath particles. The non-Hermitian asymmetry does not shift these resonances but reshapes them directionally, and it destroys the bath's Stark-induced memory through the non-Hermitian skin effect. On the other hand, we find that the impurity is localised only within an intermediate range of Stark potential. At very weak tilt strengths the bath offers no potential to trap the impurity, while at strong tilt the bath freezes into a nearly periodic charge density wave that pushes the impurity towards partial localisation. Counterintuitively, the impurity localises in the intermediate regime where the bath's initial-state memory is strongly suppressed, yet a spatially inhomogeneous density profile persists. The interaction amplifies these density fluctuations, dressing the impurity with an effective disordered potential that traps it. Since the skin effect degrades this intermediate density distribution, the impurity localisation window shifts toward stronger tilt in the non-Hermitian regime.
\end{abstract}

\maketitle
%------------------------------
% \textit{Introduction.}---
\section{Introduction}
\label{sec:intro}
Many-body localisation (MBL) is a phenomenon where an isolated interacting quantum system fails to thermalise, allowing it to retain the memory of its initial state at long times ~\cite{Anderson:1958, Nandkishore:2015, Abanin:2019}. While quenched disorder is the widely and conventionally studied way to demonstrate this, both theoretically ~\cite{Gornyi:2005, Basko:2006, Serbyn:2013, Huse:2014, Imbrie:2016, Bera:2017} and experimentally ~\cite{Schreiber:2015, Luschen:2017, Luschen:2017b}, a linear potential alone can also localise an interacting system, giving rise to the disorder-free analogue known as Stark MBL ~\cite{Schulz:2019, Nieuwenburg:2019, Taylor:2020, Guo:2021, Zeng:2023}. Its phenomenology, including logarithmic entanglement growth, level statistics, and persistent density imbalance, closely resembles that of conventional MBL, although the underlying Wannier--Stark orbitals are localised through a mechanism entirely different from that of Anderson-localised states ~\cite{Hart:1987, Bardarson:2012, Serbyn:2013b, Lukin:2019, Guo:2020}.

A related question is how a mobile particle behaves when coupled to a localised many-body system. A single impurity immersed in an Anderson-localised system can itself become localised through interactions, a proximity effect that survives even though the impurity carries no disorder of its own ~\cite{Brighi:2022, Brighi:2022b, Brighi:2023} while under different conditions the impurity can instead help thermalise the localised system ~\cite{Nandkishore:2015b, Luitz:2017, Wybo:2020}. Recently, similar questions have been explored for a Stark-localised bath as well, where a tilted lattice of bosons can slow down but not always trap a mobile impurity. The outcome depends sensitively on the structure of the bath density, with a regular charge-density wave (CDW) and a spatially disordered configuration producing qualitatively different impurity dynamics ~\cite{Falcao:2023}.

In recent years, the study of non-Hermitian systems has attracted much attention because of their fundamental properties \cite{Bender:1998, Longhi:2009, Ashida:2017, Ashida:2020, Ji:2024} and by the increasing ability to engineer effective non-Hermitian Hamiltonians in photonic, cold-atomic and other open quantum systems \cite{Xiao:2020, Hofmann:2020, WeiGou:2020, Liang:2022, Shulin:2026}. In contrast to Hermitian systems, non-Hermitian spectra can be complex and their properties depend sensitively on the boundary conditions. The well-known Hatano--Nelson model showed that the asymmetric hopping and disorder can lead to a real-to-complex spectral transition ~\cite{HatanoNelson:1996, HatanoNelson:1997}. However, under open boundary condition, no such transition occurs, instead, asymmetric hopping leads to what is known as the non-Hermitian skin effect (NHSE), in which a large number of eigenstates accumulate at one boundary. This strong boundary sensitivity has no direct analogue in conventional Hermitian lattice systems and they fundamentally modify the localisation, spectral properties, and dynamics in non-Hermitian many-body systems ~\cite{Lee:2016, Ashida:2020}. 

In view of these developments, an impurity coupled to a localised bath raises the question of how the dynamics is modified when the bath itself has additional localisation mechanism. When a linear Stark potential is added to the asymmetric hopping, the skin effect competes directly with Wannier--Stark localisation. Recent work on interacting non-Hermitian Stark chains has shown that this competition can strongly reshape the many-body phase diagram in a way that has no counterpart in non-Hermitian disordered systems ~\cite{Liu:2023, Li:2023}. 

The goal of this work is to study the dynamics of a mobile impurity coupled to a Stark localised non-Hermitian bath. Instead of focussing on the static properties previously well studied in disordered and quasi-disordered non-Hermitian systems \cite{Hamazaki:2019, Suthar:2022, Wang:2023, QianTian:2024, YalunZhang:2025, Chakrabarty:2026} and in Stark potential \cite{Li:2023, Liu:2023, QiRui:2023}, the question of impurity dynamics in such non-Hermitian systems remains to be explored. Here we address this gap by asking two questions. How does the non-Hermitian skin effect, which by itself has no analogue in the Hermitian Stark-impurity problem, modify the system dynamics and, under what conditions does the impurity localise. 

The manuscript is organised as follows. Section~\ref{sec:model} introduces the model, numerical approach, and observables. In Sec.~\ref{sec:resonance}, we discuss the impurity-mediated resonances and their effect on the bath dynamics. Section \ref{sec:skin_effect} analyses the competition between Stark localisation and the NHSE, including the associated crossover scale. In Sec.~\ref{sec:impurity}, we study impurity localisation through the bath-induced effective potential, while in Sec.~\ref{sec:entropy} we investigate the entanglement entropy to further characterise the different regimes. Finally, we summarise our findings in Sec.~\ref{sec:summary}.

\section{Model and Methods}
\label{sec:model}
% \textit{Model and methods.}---
We consider an open chain of $L$ sites with half-filled ($N_a=L/2$) hardcore bosons of species $a$ which form the bath and a single boson of species $b$ which acts as a mobile impurity. The Hamiltonian describing the system is 
\begin{equation}
    \hat H=\hat H_a+\hat H_b+\hat H_{ab}
\end{equation}
where
\begin{align}
  \hat H_a &= -J\sum_{i=1}^{L-1}\Bigl[e^{-g}\hat a_i^\dagger \hat a_{i+1}+e^{+g}\hat a_{i+1}^\dagger \hat a_i\Bigr]+\dA\sum_{i=1}^L i\,\hat n_i^a,\label{eq:Ha}\\
  \hat H_b &= -J\sum_{i=1}^{L-1}\Bigl[\hat b_i^\dagger \hat b_{i+1}+\hat b_{i+1}^\dagger \hat b_i\Bigr],\label{eq:Hb}\\
  \hat H_{ab} &= \Uab\sum_{i=1}^L \hat n_i^a \hat n_i^b.\label{eq:Hab}
\end{align}
Here, $L$ is the length of the lattice, $\hat a_i^\dagger$ ($\hat a_i$) creates (annihilates) a bath boson at site $i$, while $\hat b_i^\dagger$ ($\hat b_i$) creates (annihilates) the impurity. The corresponding number operators are $\hat n_i^a=\hat a_i^\dagger \hat a_i$ and $\hat n_i^b=\hat b_i^\dagger \hat b_i$. The parameter $J$ is the nearest neighbour hopping strength and sets the energy scale and is taken as $J=1$ throughout the paper. The bath experiences a linear Stark potential $\dA $, while the impurity is not subjected to the tilt. The parameter $g$ controls the nonreciprocity of the hopping. In contrast, the impurity hopping in Eq.~\eqref{eq:Hb} is reciprocal and therefore Hermitian. The impurity and the bath interact with strength $\Uab$ which describes the on-site density-density interaction. 
%The impurity therefore neither subjected to a Stark potential nor non-reciprocal hopping.

We study the problem using exact diagonalisation for a system of $L=10$ sites and the time evolved wavefunction $|\psi(t)\rangle$ is obtained by expanding the initial state in the eigenbasis of the Hamiltonian $\hat H$ (see Appendix \ref{appendix:dynamics}). Despite the finite-size effects present in such a small system \cite{Bera:2017, Weiner:2019}, the competition between Stark localisation and skin-effect is already clearly shown, making this a useful setting to isolate and identify its effect on the impurity dynamics. To study the dynamics of the system, we initialise the bath in the CDW configuration $|\psi_0 \rangle=|1,0,1,0, \dots\rangle$ and place the impurity at the initially unoccupied central site $i_0$.
% =L/2 +1$. 
Such a CDW state is a standard choice for probing localisation dynamics, as it provides a simple density pattern which directly measures the loss of memory of the initial state at long times. CDW state has been widely used in experiments on disorder-induced MBL ~\cite{Schreiber:2015, Luschen:2017, Luschen:2017b} and used as a diagnostic of Stark MBL as well ~\cite{Schulz:2019, Morong:2021}. We quantify the bath localisation by a quantity often used in experiments called imbalance given by \cite{Luschen:2017, Luschen:2017b, Brighi:2022}
\begin{equation}
  \mathcal{I}(t)=\frac{N_o(t)-N_e(t)}{N_o(t)+N_e(t)},
\end{equation}
where $N_o$ and $N_e$ are the bath populations of initially occupied and initially empty sites, respectively. Meanwhile, the impurity localisation is characterised by the mean-square displacement (MSD) relative to its initial position \cite{Falcao:2023},
\begin{equation}
  \mathcal{M}(t)=\sum_i(i-i_0)^2\langle n_i^b(t)\rangle-\left(\sum_i(i-i_0)\langle n_i^b(t)\rangle\right)^2,
  \label{eq:MSD}
\end{equation}
which measures the width of the impurity wavepacket after subtracting its centre-of-mass drift. This is important in the non-Hermitian system, where a narrow but drifting wavepacket could be mistaken for a delocalised state. 

Throughout the paper, we denote the averages taken over the final one-third of the evolution as $\mathcal{I}_{\rm avg}$ and $\mathcal{M}_{\rm avg}$ for imbalance and mean-square displacement respectively. We also investigate the entanglement behaviour expected from many-body localised states \cite{Abanin:2019}. We partition the chain in two halves $A$ and $B$ and calculate the half-chain von Neumann entropy 
\begin{equation}
    \mathcal{S}(t)=-\mathrm{Tr}[\rho_A(t)\ln\rho_A(t)]
\end{equation}
% $\mathcal{S}(t)=-\mathrm{Tr}[\rho_A(t)\ln\rho_A(t)]$, 
built from the reduced density matrix $\rho_A(t) =  \mathrm{Tr}_B\left[ |\psi(t)\rangle \langle\psi(t)|\right]$, where $|\psi(t)\rangle$ is the time evolved state at time $t$. Following the Refs.~\cite{Lukin:2019, Falcao:2023}, we further decompose the entropy as $\mathcal{S}=S_N+S_C$, where $S_N$ is the number entropy, which quantifies particle-number fluctuations across the bipartition, and $S_C$ is the configurational entropy within each fixed particle number sector. For each time, we first obtain the the probability $p_N(t)$ of finding a total of $N$ bath and impurity particles in subsystem $A$ by summing the diagonal elements of $\rho_A(t)$ belonging to the corresponding particle-number sector. The number entropy is $ S_N(t)=-\sum_N p_N(t)\ln p_N(t)$. The configurational entropy contribution is then obtained by subtracting the number entropy from the total entropy $ S_C(t)=\mathcal{S}(t)-S_N(t)$. This decomposition allows us to distinguish density fluctuations associated with particle tunnelling across the cut, captured by $S_N$, from the spreading of quantum information within each number sector, captured by $S_C$.
%-----------------bath localisation-------------
% \textit{Bath-imbalance resonances mediated by the impurity.}---
\section{Bath-imbalance resonances mediated by the impurity}
\label{sec:resonance}
\begin{figure}[t]
\centering
\includegraphics[width=\linewidth]{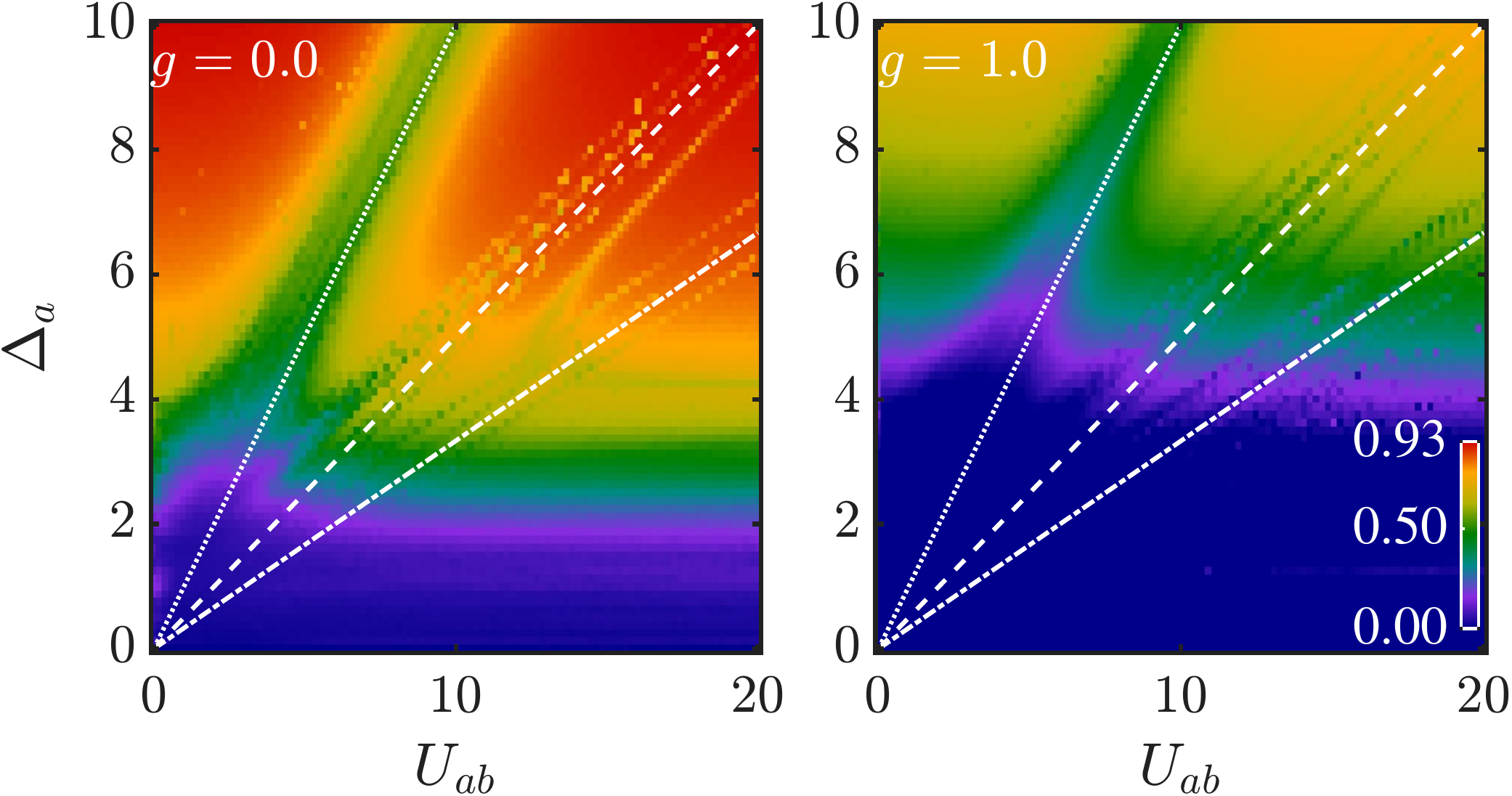}
\caption{Late-time bath imbalance $\mathcal{I}_{\rm avg}$ for the Hermitian case $g=0$ (left panel) and the non-Hermitian case $g=1$ (right panel) for a chain of $L=10$ sites at half filling ($N_a=L/2$ bath bosons), starting from the bath CDW state $|1,0,1,0,1,0,1,0,1,0\rangle$ with the impurity initially placed on the bath empty site $i_0$. Reduced-imbalance stripes are visible along $\Uab\simeq \beta \dA$, with the $\beta =1$ line (dotted) most visible and $\beta =2,3$ lines (dashed and dash-dotted, respectively) weaker. Introducing hopping asymmetry ($g=1$, right panel) reduces the imbalance as a consequence of skin effect and modifies the strength of the higher-order resonance stripes while leaving their positions unchanged.}
\label{fig:resonance}
\end{figure}

The evolution of the bath imbalance provides a direct measure of how much memory the system retains of its initial density configuration and how this memory is affected by the interplay between the Stark potential, interactions, and non-Hermitian hopping. A late-time imbalance $\mathcal{I}_{\rm avg}=1$ corresponds to perfect preservation of the initial CDW pattern, whereas $\mathcal{I}_{\rm avg}=0$ indicates a complete loss of this initial density memory. Fig.~\ref{fig:resonance} shows the average late-time bath imbalance $\mathcal{I}_{\rm avg}$ as a function of $\dA/J$ and $\Uab/J$ for the Hermitian case $g=0$ (left panel) and the non-Hermitian case $g=1.0$ (right panel). Away from specific interaction strengths $\Uab$, the bath retains a large imbalance, while significant dips appear along the lines
\begin{equation}
\Uab\simeq \beta\dA,\qquad \beta=1,2,\dots.
\label{eq:resonance}
\end{equation}
The leading ($\beta=1$) dip can be understood by a resonance condition between two configurations that differ by a single hop of the bath-particle. Consider a scenario when the impurity is at site $i$ and a bath boson is at its right neighbour $i+1$. The bath particle has energy $\dA(i+1)$, whereas a bath boson occupying the impurity site has energy $\dA i+\Uab$, including the interaction cost of occupying the same site as the impurity. A hop towards the left site $i+1\rightarrow i$ is therefore resonant when $\dA(i+1)=\dA i+\Uab$, giving $\Uab=\dA$. On the other hand, a bath boson initially on the left neighbour $i-1$ cannot resonantly hop onto the impurity for positive $\Uab$ and $\dA$. Thus, the Stark potential selects the right side of the impurity as the resonant channel. In this sense, the impurity provides a local pathway for resonant tunnelling of bath particles thereby reducing the memory of their initial configuration.

However, the resonance condition does not determine the directionality of the corresponding hopping amplitudes. The two resonant configurations are connected by opposite hopping mechanisms. Once a bath particle occupies the impurity site, it can leave through $\beta$ consecutive rightward hops which satisfies the resonance condition. In contrast, the reverse process brings a bath particle initially located $\beta$ sites to the right back onto the impurity through $\beta$ consecutive leftward hops. For the bath with $J/\Delta_a\ll 1$, these two processes have different amplitudes,
\begin{equation}
\Tbeta(+)\sim\frac{J^\beta e^{\beta g}}{(\beta -1)!\,\dA^{\beta -1}},\qquad
\Tbeta(-)\sim\frac{J^\beta e^{-\beta g}}{(\beta -1)!\,\dA^{\beta -1}},
\label{eq:Tn}
\end{equation}
where $\Tbeta(+)$ denotes the rightward hop from the impurity to a site $\beta$ lattice to its right, while $\Tbeta(-)$ denotes the reverse leftward tunnelling. Importantly, these are two matrix elements connecting the same pair of configurations and therefore obey the same energy-matching condition, $\Uab\simeq \beta\dA$.
 
Higher-order resonances at $\Uab\simeq \beta\dA$ arise from $\beta$-step hopping processes connecting positions in which the bath particle occupies the impurity site to one in which it is displaced by $\beta$ sites. The total Stark-energy difference between these configurations is $\beta\dA$, which is compensated by the interaction energy $\Uab$ when the resonance condition is met. The intermediate states are separated from the resonance condition by successive Stark-energy offsets. Consequently, the amplitudes scale as in Eq.~\eqref{eq:Tn}, with  $\Tbeta^2(\pm) \propto \left(\frac{J}{\dA}\right)^{2(\beta -1)}e^{\pm 2\beta g}$. Thus, higher-order processes become weaker with increasing order due to the Stark potential, while the hopping asymmetry introduces an increasingly strong directional imbalance between the two tunnelling channels.

For $g>0$, each rightward tunnelling process is enhanced by a factor of $e^{g}$, whereas leftward tunnelling is suppressed by $e^{-g}$. This directional asymmetry enhances the resonant dynamics towards the right and leads to a stronger reduction of the bath imbalance, as shown in the right panel of Fig.~\ref{fig:resonance}. In addition to the pronounced resonant dips, the overall late-time average imbalance is substantially reduced for $g=1$ compared with the Hermitian case $g=0$. This indicates that the hopping asymmetry weakens the bath's ability to retain its initial CDW memory. This directional dynamics follows the same direction as that of the non-Hermitian skin effect, discussed in the next section. Importantly, the hopping asymmetry does not shift the resonance positions, which are determined by the Stark and interaction energies.

\section{Effect of hopping asymmetry on the bath}
% \textit{Skin effect versus Stark localisation in the bath.}---
\label{sec:skin_effect}
% Since $\hat H_a$ has only single particle operators, its $N_a$-boson eigenstates factorise into single-particle orbitals of $h_a$, the one-body Hamiltonian corresponding to Eq.~\eqref{eq:Ha} (see Appendix \ref{appendix:bloch_osc}). 
Since $\hat H_a$ has only single-particle boson operators with hardcore constraint, the bath can be mapped onto a system of non-interacting fermions in one dimension through the Jordan--Wigner transformation. The resulting
many-body eigenstates are therefore Slater determinants constructed
from the single-particle eigenstates of $h_a$, the one-body Hamiltonian
corresponding to Eq.~\eqref{eq:Ha} (see Appendix~\ref{appendix:bloch_osc}). Under open boundary conditions, the asymmetric hopping in $h_a$ can be removed by a similarity transformation ~\cite{Hatano:2021}. Writing the right eigenstate as $\psi^R(n)=e^{gn} \varphi(n)$, the right eigenvalue equation becomes the Hermitian Wannier--Stark equation \cite{Wannier:1962, Fukuyama:1973, Hart:1987} 
\begin{equation}
-J\varphi(n+1) -J\varphi(n-1)+\dA n\,\varphi(n)=E\varphi(n).
\end{equation}
For an open chain the spectrum of $h_a$ is independent of $g$ and is the same as that of the Hermitian Stark case, while the right and left eigenstates acquire  exponential envelopes with opposite signs ~\cite{Hart:1987, HatanoNelson:1996, Hatano:1998, Xie:2024},
\begin{equation}
\psi_\nu^R(n)\approx e^{gn}\mathcal{J}_{n-\nu}\left(\frac{2J}{\dA}\right),\qquad
\psi_\nu^L(n)\approx e^{-gn}\mathcal{J}_{n-\nu}\left(\frac{2J}{\dA}\right).
\label{eq:eigenstates}
\end{equation}
Here $\mathcal{J}_{n-\nu}(x)$ denotes the Bessel function of the first kind, and $\mathcal{J}_{n-\nu}(\frac{2J}{\dA})$ gives the Wannier--Stark orbital with ladder energies $E_\nu=\dA \nu$ and a characteristic Stark localisation length $\ell_S\sim2J/\dA$ ~\cite{Wannier:1962, Fukuyama:1973, Hart:1987}. For $g>0$, the right eigenstates are shifted towards the right boundary, whereas the left eigenstates are accumulated towards the left boundary. Thus the right and the left eigenstates have different spatial profiles which is expected from a non-Hermitian system. For this reason, the localisation properties and dynamics are discussed using no-jump condition (see Appendix \ref{appendix:dynamics}) \cite{Hatano:1998, Ashida:2020} . In this system, the right-eigenstate density contains the factor $e^{2gn}$ and therefore has a skin length $\ell_{\rm skin}=1/(2g)$.

\begin{figure}[t]
    \centering
    \includegraphics[width=\linewidth]{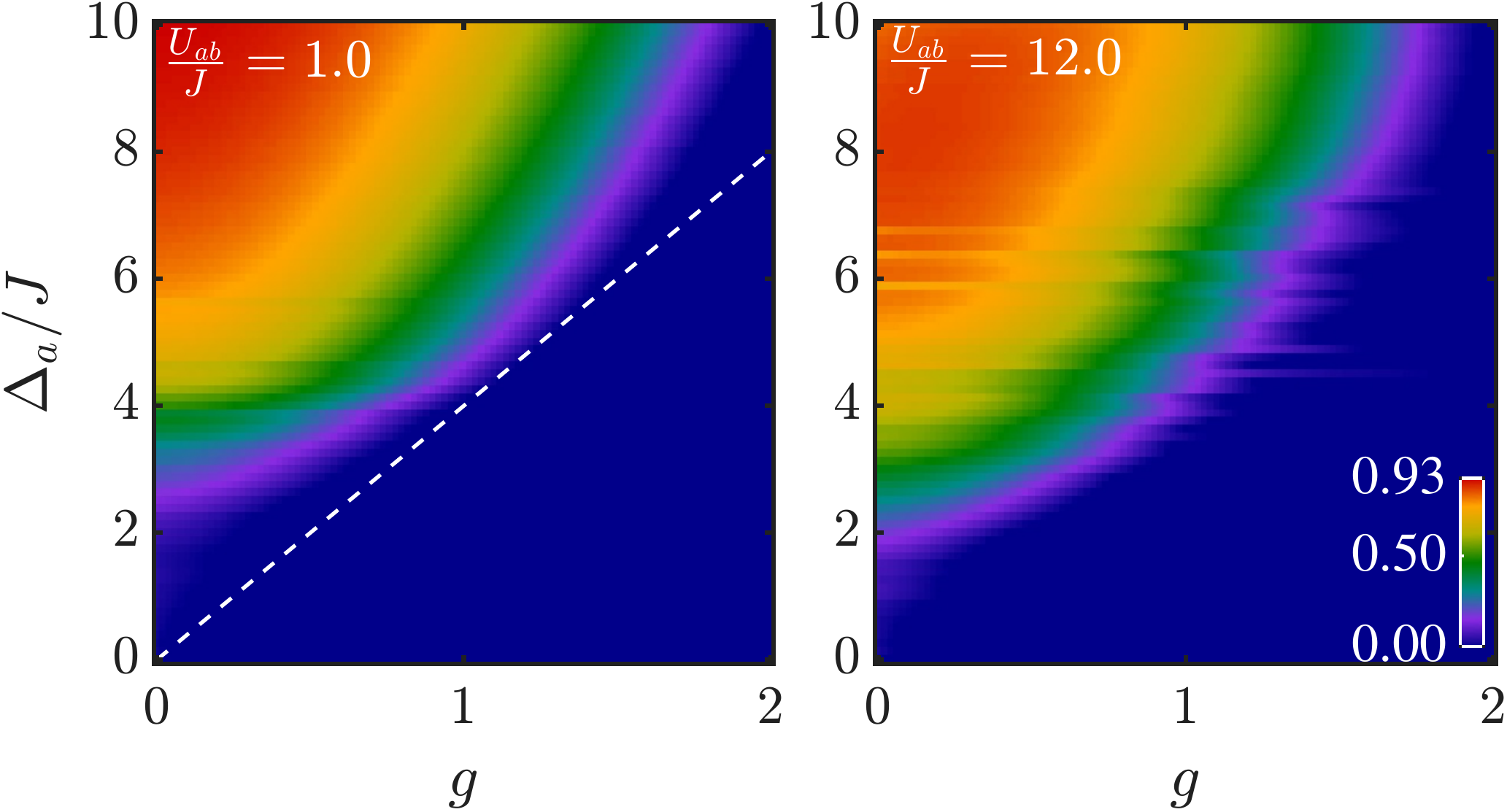}
    \caption{Late-time imbalance $\mathcal{I}_{\rm avg}$ as a function of hopping asymmetry $g$ and tilt $\dA/J$, for weak impurity-bath interaction $\Uab/J=1$ (left panel) and strong interaction $\Uab/J=12$ (right panel), for the half-filled CDW initial state as in Fig.~\ref{fig:resonance}. In both panels the upper-left region (small $g$, large $\dA/J$) indicates a robustly Stark-localised bath with $\mathcal{I}_{\rm avg}$ close to unity, while the lower-right region (large $g$, small $\dA/J$) shows that the CDW memory is lost due to the skin effect. The white dashed line shows $g_c$ of the Eq. \eqref{eq:gc} for the non-interacting case.  Large interaction between the impurity and the bath ($\Uab/J=12.0$) additionally shows fine-scale, non-monotonic structure across $\dA/J\approx4$--$7$ that is absent in the left panel at the same $g$. This structure coincides with the $\beta=3$ ($\dA/J=4$) and $\beta=2$ ($\dA/J=6$) resonance lines of Eq.~\eqref{eq:resonance}, and the imbalance is modified by $g$ for strong interaction.}
    \label{fig:skin_effect}
\end{figure}

For the single-particle propagator, the same similarity transformation gives
\begin{align}
G(i,k,t)=\langle i|e^{-ih_at}|k\rangle
&=\sum_\nu \frac{\langle i|\psi_\nu^R\rangle \langle\psi_\nu^L|k\rangle} {\langle\psi_\nu^L|\psi_\nu^R\rangle}
e^{-iE_\nu t} \nonumber \\ 
&=e^{g(i-k)}G_{\rm H}(i,k,t)\;,
\end{align}
where $G_{\rm H}$ is the propagator of the Hermitian Wannier--Stark case. Thus, for the open-chain single-particle case, the non-Hermitian dynamics differs from the Hermitian Stark dynamics by an exponential factor $e^{g(i-k)}$. This is a direct consequence of the similarity transformation ~\cite{Hatano:1998}, which enhances propagation towards the right of the chain while suppressing propagation towards the left. The extension to the many-body bath is less straightforward, as the non-unitary evolution modifies the spatial distribution of the initially occupied orbitals. For $g>0$, the asymmetric hopping shifts the bath density towards the right boundary, while suppressing it on the left. As a result, the contrast between the initially occupied and empty sites is reduced, leading to a decrease in the late-time imbalance. This behaviour is shown in Fig.~\ref{fig:skin_effect}, where increasing $g$ at fixed $\dA$ suppresses the average imbalance $\mathcal{I}_{\rm avg}$.

The competition between the two localisation mechanisms can be established by comparing their characteristic length scales. When $\ell_{\rm skin}\gg\ell_S$, the Stark-localised orbital is confined to a region where the skin envelope changes only weakly, so the spatial profile is mainly determined by the Stark localisation. In the opposite limit, $\ell_{\rm skin}\ll\ell_S$, the skin envelope varies strongly across the Wannier--Stark orbital and significantly shifts the right eigenstate towards the boundary. Equating the two scales gives 
\begin{equation}
g_c(\dA)\approx\frac{\dA}{4J},
\label{eq:gc}
\end{equation}
indicating that a larger Stark tilt is required to compensate a larger hopping asymmetry before the skin effect becomes dominant. This single-particle estimate captures the overall crossover reasonably well at weak impurity coupling as shown in the left panel of Fig.~\ref{fig:skin_effect} where the white dashed line corresponds to Eq.\eqref{eq:gc}. At strong coupling, $\Uab/J=12.0$ however, the behaviour is no longer described by the single particle picture. The decay of $\mathcal{I}_{\rm avg}$ with $g$ is no longer smooth and it develops visible fine-structure changes in the range $\dA/J\approx4-7$ as shown in the right panel of Fig.~\ref{fig:skin_effect}. For $\Uab/J=12.0$, this range contains the $\beta=3$ ($\dA/J=4$) and $\beta=2$ ($\dA/J=6$) impurity-mediated resonances of Eq. ~\eqref{eq:resonance}. This structure is absent from the weakly-interacting panel at the same $g$ indicating that this arises from the impurity mediated resonant channel. Moreover, the strength of these resonant processes is modified by $g$, which is visible in Fig.~\ref{fig:skin_effect}.  Thus, in the strongly interacting regime, the skin effect and impurity-mediated resonances are coupled rather than acting as independent mechanisms.

% \textit{Impurity localisation through interaction.}---
\section{Impurity localisation through interaction}
\label{sec:impurity}

The impurity described by the Hamiltonian in Eq.~\eqref{eq:Hb} has no intrinsic mechanism for localisation and therefore any changes in the free motion of the impurity can be attributed to the interaction with the bath through $U_{ab}$. The density-density interaction allows the bath to act as a time-dependent effective potential
\begin{equation}
V_{\rm eff}(i,t)\approx\Uab\langle n_i^a(t)\rangle \;,
 \label{eq:Veff}
\end{equation}

\begin{figure*}
    \centering
    \includegraphics[width=\linewidth]{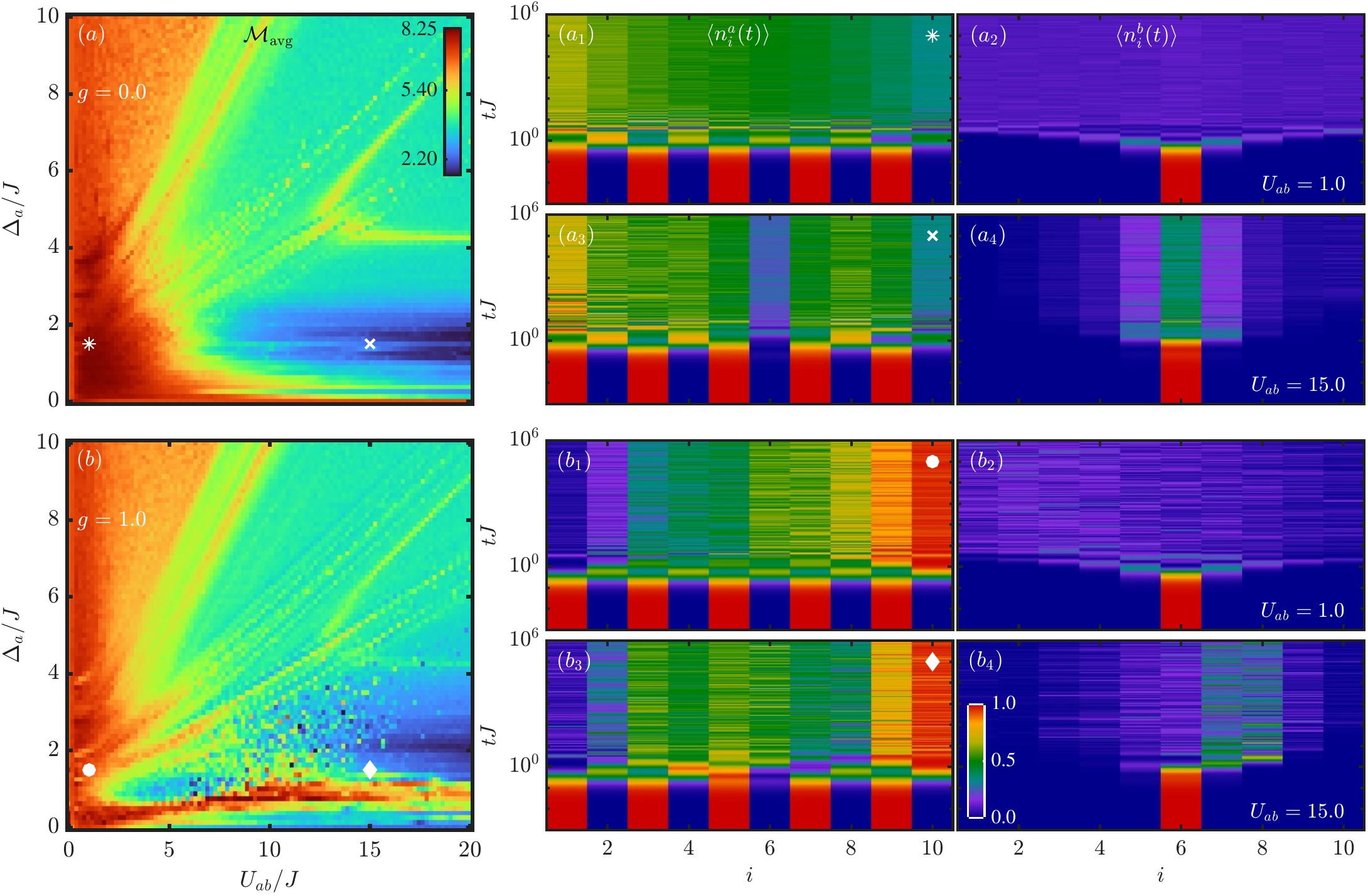}
    \caption{Impurity localisation induced by the bath through the effective potential $V_{\rm eff}$ in Eq.~\eqref{eq:Veff}. Late-time impurity mean-square displacement $\mathcal{M}_{\rm avg}$ as a function of the bath tilt $\dA/J$ and impurity-bath interaction $\Uab/J$ for the Hermitian case $g=0$, $(a)$ and the non-Hermitian case $g=1$, $(b)$. For $g=0$, the strongest localisation occurs around $\dA/J\sim1.5$. For $g=1$, the skin effect shifts the localisation region towards larger $\dA/J$. $(a_1)$--$(a_4)$ show time evolution of the bath density $\langle n_i^a(t)\rangle$ and impurity density $\langle n_i^b(t)\rangle$ at $\dA/J=1.5$ and $g=0$. For $\Uab/J=1.0$ $(a_1)$,$(a_2)$, the bath loses its initial CDW structure while retaining weak signatures of the Stark potential at late times and the impurity spreads across the lattice $(a_2)$. For $\Uab/J=15.0$, $(a_3)$,$(a_4)$, the bath evolution is qualitatively similar $(a_3)$, but the impurity remains confined near its initial site $(a_4)$, showing that impurity localisation is controlled by the spatial structure of the bath density rather than by CDW memory alone. $(b_1)$--$(b_4)$ show the bath and impurity densities for $g=1$. At $\Uab/J=1.0$ $(b_1)$,$(b_2)$, the skin effect shifts the bath density $(b_1)$ towards the right boundary, while the weakly coupled impurity remains delocalised $(b_2)$. At $\Uab/J=15.0$ $(b_3)$,$(b_4)$, the bath again develops a right-biased density profile $(b_3)$, but the impurity does not disperse but instead localises with a rightward bias due to the interaction with the skin affected bath $(b_4)$.}
    \label{fig:MSD}
\end{figure*}

\noindent which shows that the localisation of the impurity is determined by the spatial and time dependence of the density distribution of the bath. Importantly, this means that the impurity's localisation is not determined solely by the retention of the initial CDW pattern of the bath. 

For the $L=10$ system considered here, the impurity dynamics involves only a finite number of discrete single-particle levels. Even in the decoupled limit $\Uab/J=0$, $\mathcal{M}(t)$ therefore does not converge to a stationary value. The freely evolving impurity remains a superposition of discrete eigenmodes, and its density profile continues to oscillate rather than relaxing to a stationary uniform distribution \cite{Bera:2017, Weiner:2019}. Since the impurity is completely decoupled from the bath in this limit, its dynamics is independent of both $\dA$ and $g$. We therefore use the same late-time averaging procedure as in the interacting system and define the free impurity reference as $\mathcal{M}_{\rm free} \equiv \mathcal{M}_{\rm avg} \big|_{\Uab=0} \simeq6.8$. For comparison, a uniform density distribution gives the value $\mathcal{M}_{\rm unif}
= \frac{1}{L}\sum_{i}(i-i_0)^2 -\left[\frac{1}{L}\sum_{i}(i-i_0) \right]^2 =8.25$ which is close to the largest $\mathcal{M}_{\rm avg}$ values over the parameter range for the chosen initial state. The lower value of $\mathcal{M}_{\rm free}$ reflects the finite size dynamics of the decoupled impurity and the resulting time averaged density profile rather than a uniform stationary distribution. We therefore use $\mathcal{M}_{\rm free}$ as the reference for identifying impurity localisation. We take $\mathcal{M}_{\rm avg}<\mathcal{M}_{\rm{free}}/3\simeq2.2$ as a signature for strong impurity localisation, while $\mathcal{M}_{\rm avg} \gtrsim 0.8 \mathcal{M}_{\rm{free}} \simeq5.4$ corresponds to delocalisation. The remaining region is interpreted as weakly or partially localised. 

Fig.~\ref{fig:MSD}$(a)$ shows the late-time averaged MSD as a function of $\dA$ and $\Uab$. The impurity remains delocalised at weak interactions and small tilts, with $\mathcal{M}_{\rm avg}$ close to its largest values. In this regime, the bath rapidly loses memory of its initial CDW state, and increasing $\dA$ has little effect on the impurity dynamics. Similarly, increasing $\Uab$ does not significantly confine the impurity, as the bath itself retains only a small imbalance in this regime. 

A different behaviour appears at intermediate tilt. Though the bath does not retain its initial CDW memory, its density still develops a spatial structure that can confine the impurity through $V_{\rm eff}$ and results in the suppression of the impurity MSD indicating localisation of the impurity. To illustrate this, we show the bath density $\langle n_i^a(t)\rangle$ and the impurity density $\langle n_i^b(t)\rangle$ as a function of time at fixed $\dA/J=1.5$ and for $\Uab/J=1.0$ and $\Uab/J=15.0$. For $\Uab/J=1.0$, the bath density loses its initial structure, with small effects of Stark effect remaining at late times as shown in Fig.~\ref{fig:MSD}$(a_1)$. Impurity density on the other hand, completely spreads out across the lattice indicating that the effective potential is too weak to localise it as shown in Fig.~\ref{fig:MSD}$(a_2)$.

At the same tilt but stronger interaction $\Uab/J=15.0$, the bath density shows a similar evolution as shown in Fig.~\ref{fig:MSD}$(a_3)$, yet the impurity becomes confined as illustrated in Fig.~\ref{fig:MSD}$(a_4)$. Its density remains largely confined around the initial site with less spreading in contrast to $\Uab/J=1.0$ case. This suggests that the localisation is not just a consequence of the bath retaining the CDW memory, rather, it results from the interaction amplifying the spatial density distribution in the bath. This process is similar to interaction-induced impurity localisation in a disordered bath ~\cite{Brighi:2022, Brighi:2022b, Brighi:2023}. However, here the effective trapping potential is generated dynamically by the Stark-localised bath rather than by externally imposed quenched disorder. 

The large-tilt regime provides a contrasting limit to the intermediate regime. For $\dA\gg J$, bath motion is strongly suppressed and its density remains close to its initial CDW pattern. The impurity then sees an approximately static period-two potential, $V_{\rm eff}(i) \simeq \Uab \langle n_i^a\rangle \approx \Uab \sum_m \delta_{i,2m+1}$ which is the lattice analogue of a Kronig--Penney potential. In an infinite system, such a periodic potential modifies the impurity motion but does not by itself localise the particle, so extended Bloch-like motion is expected as discussed in Ref.~\cite{Falcao:2023}. Our results only partially show this behaviour. For $\dA/J\gtrsim5.0$ and $\Uab/J\gtrsim10.0$, $\mathcal{M}_{\rm avg}$ saturates around $4.0$. This is below the free value $\mathcal{M}_{\rm{free}}$, but above the values in the localisation region. We attribute this mainly to the small system size, with $L=10$, the period-$2$ potential contains only five unit cells, insufficient for a clear band structure or fully developed Bloch-like states. The observed saturation may therefore be a finite-size effect.
% , and the study of larger systems are required to determine whether Bloch-like scattering states appear in this regime.

When the hopping asymmetry $g=1.0$ is introduced, the weak-tilt and weak-interaction regime remains essentially unchanged as shown in Fig.~\ref{fig:MSD}$(b)$. For $\dA/J \lesssim1.0$ and $\Uab/J \lesssim 3.0$, the $\mathcal{M}_{\rm avg}$ remains close to its $g=0$ counterpart and well above the delocalisation threshold.

The main change occurs in the intermediate tilt regime, where the impurity is most strongly localised for Hermitian case $g=0$. At $\dA/J\simeq1.5$ and $\Uab/J=1.0$, the asymmetric hopping modifies the bath density distribution and weakens the spatial structure of the effective potential seen by the impurity. As a result, $\mathcal{M}_{\rm avg}$ increases towards the delocalised region. As can be seen from the evolution of the bath density in Fig.~\ref{fig:MSD}$(b_1)$, the bath density now is biased towards the right and the density accumulates at the right boundary. The impurity on the other hand is not affected by the hopping asymmetry as $\Uab/J$ is too weak to have any effect on the impurity and therefore the impurity spreads across the lattice as shown in Fig.~\ref{fig:MSD}$(b_2)$.

At the same tilt but for $\Uab/J=15.0$, the bath density develops a bias towards the right edge due to the skin effect as shown in Fig.~\ref{fig:MSD}$(b_3)$ and has qualitatively similar density profile as that of $\Uab/J=1.0$ case. In contrast to the Hermitian case, 
where the impurity is strongly confined to its initial site, the impurity now spreads predominantly towards the right of the lattice as shown in Fig.~\ref{fig:MSD}$(b_4)$. The large interaction therefore does not restore the localisation found at $g=0$. Instead, the asymmetric bath changes the spatial structure of the effective potential strongly enough to overcome the interaction-induced confinement.

The localisation window is not, however, completely destroyed. Instead, it is shifted towards larger tilts. For $\dA/J \gtrsim1.5$, the $\mathcal{M}_{\rm avg}$ decreases again to the values around the localisation threshold (see Fig.~\ref{fig:MSD}$(b)$). A stronger Stark potential is therefore required to compensate for the delocalising effect of the non-Hermitian skin effect and to maintain sufficient bath density memory to trap the impurity. The competition between Stark localisation and the skin effect thus shifts the interaction-induced impurity-localisation window to larger $\dA$. 

We also observe that $\mathcal{M}_{\rm avg}$ shows narrow features along the resonant lines $\Uab\simeq\beta\dA$. Qualitatively, at these interaction strengths, the impurity-mediated resonance enhances bath tunnelling and partially melts the CDW pattern lowering the average imbalance $\mathcal{I}_{\rm avg}$. This resulting density redistribution modifies the effective potential in Eq.~\eqref{eq:Veff}, that enhances the impurity motion dynamically. The narrow lines in $\mathcal{M}_{\rm avg}$ thus indicate that the bath-imbalance resonances that destabilise the CDW pattern can also weaken the interaction-induced impurity localisation.

% \textit{Entanglement dynamics.}---
\section{Entanglement dynamics}
\label{sec:entropy}

The half-chain entanglement entropy provides a complementary probe for the impurity--bath dynamics, beyond the measures of average imbalance and impurity's average MSD. We write the entropy as $\mathcal{S}=S_N+S_C$, as mentioned in Sec.~\ref{sec:model}. The contribution $S_N$ is sensitive to particle transport across the cut. This motion broadens the particle number distribution and increases $S_N$, while accumulation away from the cut can suppress it. Thus, $S_N$ mainly reflects particle transport across the bipartition. In contrast, $S_C$ characterises the mixing of different configurations within each particle number sector. A large $S_C$ therefore indicates that a wider range of bath--impurity states are involved in the dynamics, providing a measure of the bath particles that dress the impurity.

\begin{figure*}[htbp!]
\centering
\includegraphics[width=\linewidth]{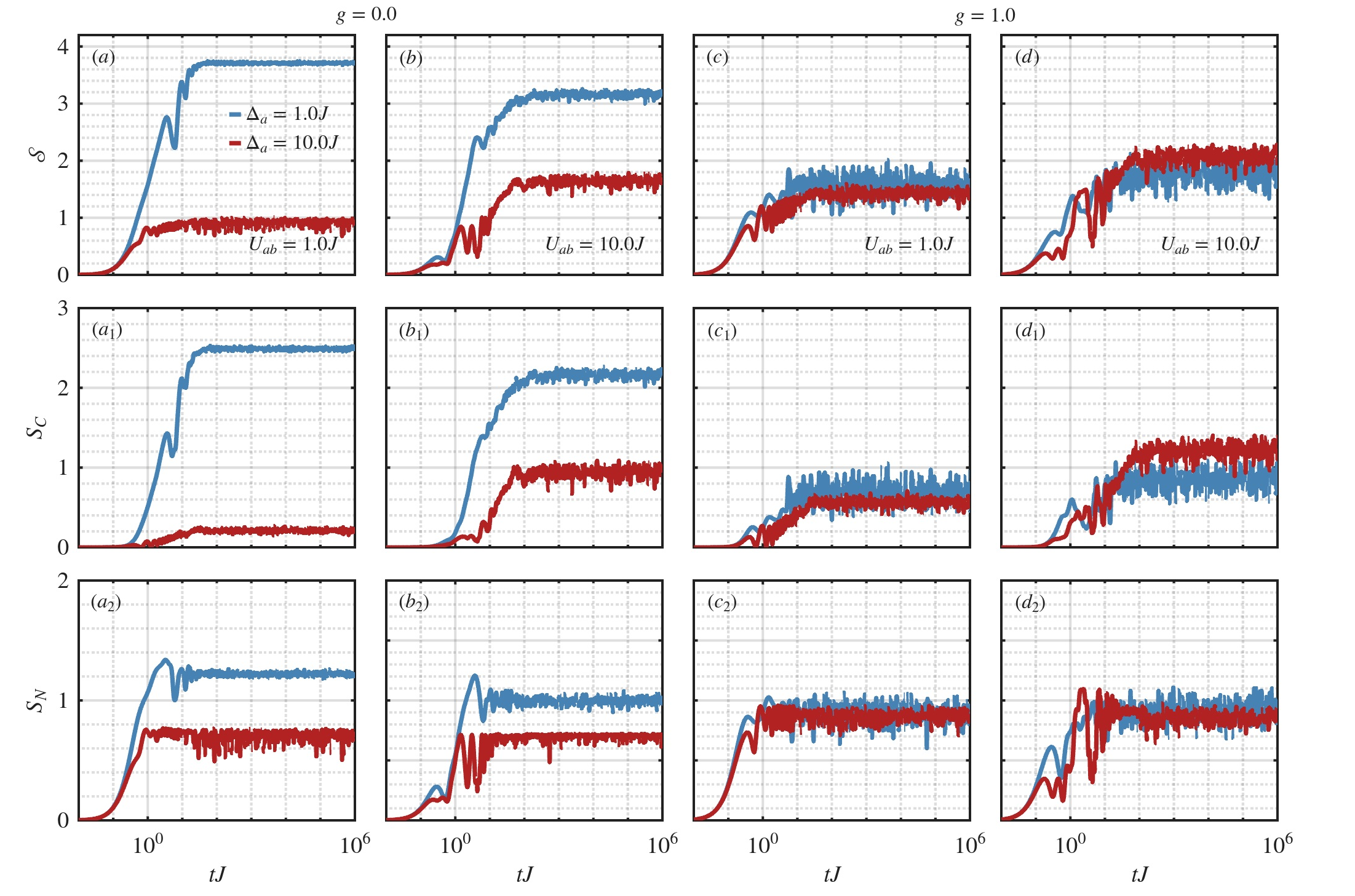}
\caption{Entanglement dynamics for weak coupling $\Uab/J=1.0$ in the Hermitian case, $(a)$--$(a_2)$, and in the presence of the skin effect, $(c)$--$(c_2)$, and for strong coupling $\Uab/J=10.0$ in the Hermitian case, $(b)$--$(b_2)$, and with the skin effect, $(d)$--$(d_2)$. The first, second, and third panels in each case show the total entropy $\mathcal{S}$, configurational entropy $S_C$, and particle-number entropy $S_N$, respectively. Blue and red curves correspond to $\dA/J=1$ and $\dA/J=10.0$. At weak coupling, increasing the tilt suppresses the total entropy and changes its dominant contribution from $S_C$ to $S_N$, consistent with Stark localisation. The skin effect further suppresses the entropy at weak tilt by reducing the configurational mixing. At $\Delta_a/J=10.0$, the resonant case $U_{ab}/J=10.0$ exhibits larger $S$ and $S_C$ than the case $U_{ab}/J=1.0$. Increasing $g$ enhances the entropy at the resonant point, while a clear resonance signature remains. The skin effect modifies both contributions and reduces the distinction between the weak and strong tilt cases.}
\label{fig:entropy}
\end{figure*}

The total half-chain von Neumann entanglement entropy with the decomposition is shown in Fig.~\ref{fig:entropy}. We first consider the Hermitian case, $g=0$, starting with weak coupling $\Uab/J=1.0$. For weak tilt $\dA/J=1.0$ shown as the blue curve in Fig.~\ref{fig:entropy}$(a)$, the system  generates entanglement rapidly, with $\mathcal{S}$ saturating around $\approx3.7$. The configurational part accounts for more than half of the total entropy (see panel Fig. \ref{fig:entropy}$(a_1)$), showing that many different bath--impurity states are mixed in this regime. Increasing the tilt to $\dA/J=10.0$ shown as the red curve in Fig.~\ref{fig:entropy}$(a)$, strongly reduces the entropy with $\mathcal{S}$ saturating at $\approx0.92$, nearly a factor of four smaller than the weak tilt. The dominant contribution also changes from $S_C$ to $S_N$ (see panel Fig.~\ref{fig:entropy}$(a_2)$). Thus the strong tilt suppresses configurational mixing and bath rearrangements, while particle number fluctuations across the cut become the main contributors to the remaining entanglement consistent with a Stark localised bath. 

We next increase the interaction to $\Uab/J=10.0$, shown in the second column of Fig.~\ref{fig:entropy}. At weak tilt, $\dA/J=1.0$, the total entropy saturates around $\approx3.2$, with $S_C$ again becoming the dominant contributor. Thus, increasing the interaction from $\Uab/J=1.0$ to $\Uab/J=10.0$ at weak tilt does not change the qualitative character of the entanglement, but maintains substantial configurational mixing through the stronger impurity--bath coupling. At strong tilt $\dA/J=10.0$, the behaviour is different. These values of the parameters correspond to the $\beta=1$ impurity-mediated tunnelling resonance, where $\Uab=\dA$. The total entropy increases to $\mathcal{S}\approx 1.65$, substantially above the off-resonant value $\mathcal{S}\approx 0.92$ observed for $\Uab/J=1.0$. This increase can be mainly attributed to $S_C$, while $S_N$ remains closer to its weak coupling value. The resonance therefore opens additional bath--impurity configurations and enhances $S_C$. This is consistent with the resonance observed in the imbalance in Fig.~\ref{fig:resonance}, where the impurity-mediated process allows the bath to rearrange through the impurity leading to decrease in the imbalance value.

The comparison between the two coupling strengths at $g=0$ therefore shows that the effect of increasing $\Uab$ depends strongly on the tilt. At weak tilt, both $\Uab/J=1.0$ and $\Uab/J=10.0$ generate substantial configurational entanglement, with only a moderate change in the total entropy. At strong tilt, however, increasing $\Uab$ brings the system to the $\beta=1$ resonance condition and substantially increases the configurational contribution. The entropy therefore follows the same competition between Stark localisation and interaction-induced bath rearrangement seen in the imbalance dynamics.

We now consider  the effect of the asymmetric hopping $g=1.0$, shown in the two right columns of Fig.~\ref{fig:entropy}. For weak coupling $\Uab/J=1.0$, the weak tilt case $\dA/J=1.0$ gives $\mathcal{S}\approx1.57$, considerably smaller than the corresponding Hermitian value of $\approx3.7$  (see Figs.~\ref{fig:entropy}$(a)$ and $(c)$). The reduction mainly comes from the contribution from $S_C$ (Fig.~\ref{fig:entropy}$(c_1)$), indicating that the skin effect limits the mixing of bath--impurity states during the dynamics. Increasing the tilt to $\dA/J=10.0$ shows a different behaviour. In contrast to the strong suppression seen when going from $\dA/J=1.0$ to $\dA/J=10.0$ for the Hermitian case $g=0$, the total entropy now in the non-Hermitian regime remains of comparable magnitude and saturates around $\approx1.43$. The configurational contribution is also increased compared to the Hermitian case (see Figs.~\ref{fig:entropy}$(a_1)$ and $(c_1)$), while $S_N$ changes only moderately. Thus, in the presence of the skin effect, the strong tilt no longer produces the clear suppression of configurational mixing seen in the Hermitian case. Instead, the non-reciprocal hopping partially restores the configurational mixing that is suppressed by the Stark potential.

For strong coupling, $\Uab/J=10.0$, the same comparison between weak and strong tilt differs considerably with the Hermitian case. At weak tilt, $\dA/J=1.0$, the total entropy is reduced relative to the corresponding $g=0$ case (Fig.~\ref{fig:entropy}$(d)$). The main reduction occurs in $S_C$, while $S_N$ remains of similar value. Thus, despite the strong impurity--bath interaction, the skin effect reduces the configurational mixing generated during the dynamics. At strong tilt, $\dA/J=10.0$, the system is at the $\beta=1$ resonance. Here $\mathcal{S}$ increases to $\approx2.09$, compared with $\approx1.65$ at $g=0$. Both $S_C$ and $S_N$ contribute to this increase, while their relative contributions remain approximately unchanged (see panels $(d_1), (d_2)$ in Fig.\ref{fig:entropy}). The increase of the total entropy is consistent with the increase of the resonant tunnelling amplitude, $\mathcal{T}_\beta\propto e^{\beta g}$, which increases the impurity-mediated bath rearrangement.

Comparing the two coupling strengths $\Uab/J=1.0$ and $\Uab/J=10.0$ at $g=1$ further shows that the distinction between weak and strong coupling becomes less clear than in the Hermitian case. At weak tilt, increasing $\Uab$ changes the total entropy only moderately, while the configurational contribution remains strongly affected by the skin-induced suppression of bath rearrangements. At strong tilt, the resonant condition at $\Uab/J=10.0$ remains visible, but the entropy difference between the weak and strong coupling cases is smaller than in the Hermitian system. Thus, although the resonance still leaves a clear signature in the strong-coupling case, the overall entropy is strongly modified by the non-Hermitian dynamics. 

Thus, the entropy decomposition distinguishes the weak and strong tilt regimes much more clearly for $g=0$ than for $g=1$. In the non-Hermitian case, the interplay of Stark localisation, impurity--bath interaction, and non-reciprocal hopping modifies both entropy contributions, making the different dynamical regimes less distinct.

% \textit{Conclusion.}
\section{Summary}
\label{sec:summary}

In this study we investigated the dynamics of a clean mobile impurity coupled to a Stark-localised bath with non-reciprocal hopping. We observed resonant tunnelling of bath particles at $\Uab\simeq\beta\dA$ changes the initial CDW state and decreases the imbalance and changes the configurational entanglement entropy according to the hopping asymmetry $g$. The resonance positions are not shifted by the non-Hermitian asymmetry, while the corresponding tunnelling amplitudes become strongly directional, scaling as $\mathcal{T}_\beta\propto e^{\pm\beta g}$. Thus, the impurity is not simply moving in a fixed localised background. Its motion and the bath are in constant feedback and modify the environment that controls impurity's dynamics.

The impurity localisation itself is determined by the density structure retained by the bath rather than by the degree of Stark localisation alone. At weak tilt, the bath quickly loses its initial CDW memory and provides little spatial structure for localising the impurity. At strong tilt, the bath remains close to the initial CDW and produces an effective periodic potential, which also does not efficiently localise the impurity. The strongest localisation occurs between these two limits, where the bath retains essentially no memory of its initial state while still undergoing sufficient rearrangements to generate a spatially inhomogeneous effective potential to trap the impurity.

The non-Hermitian skin effect changes this balance by driving the bath density towards one boundary and modifying the CDW structure. This weakens the bath-induced potential responsible for impurity localisation and shifts the localisation regime towards stronger Stark potentials. At the same time, the skin effect can either suppress or enhance the dynamics depending on the regime. Away from resonance, it can reduce the configurational entropy associated with impurity dressing, while in the Stark-localised regime it can partially restore configurational entropy that is suppressed by the strong tilt. The resonant channel is enhanced in the same direction as the non-reciprocal tunnelling amplitude, leading to stronger bath rearrangements at $\Uab\simeq\beta\dA$.

In the Hermitian case, the $S_C$ and $S_N$ contributions show a clear distinction between the different tilt and interaction regimes. With increasing non-Hermiticity, however, the skin effect modifies both contributions and reduces this distinction. The entanglement therefore reflects not only the localisation of the bath and impurity, but also the redistribution of particles and configurations caused by the non-reciprocal dynamics.

This study may provide a route to controlling impurity localisation without introducing static disorder and shows how non-Hermitian asymmetry can be used to tune interaction-driven localisation in disorder-free many-body systems.

\section{Acknowledgements}
MSH thanks SR Padhi and Salassal for fruitful discussions and insights on the entropy. The authors acknowledge support from the National Science and Technology Council (NSTC), Taiwan, under the Grants No. NSTC-115-2112-M-001-035-MY3 and No. NSTC-115-2119-M-001-009, and from Academia Sinica under Grant AS-CDA-113-M04 and for support from TG 1.2 of NCTS at NTU. J.-S.Y. acknowledges support from the National Science and Technology Council (NSTC), Taiwan, under Grant No. NSTC 115-2112-M-003-020, from Higher Education Sprout Project of National Taiwan Normal University and the Ministry of Education (MOE), Taiwan, and from TG 3.2 of NCTS.

%--------------------Appendix--------------
\appendix

\section{Dynamics of non-Hermitian systems}
\label{appendix:dynamics}

Under the no-jump condition, the time evolution of the many-body state is given by \cite{Daley:2014, Brody:2014}
\begin{equation}
|\psi(t)\rangle =
\frac{e^{-i\hat Ht}|\psi_0\rangle}
{\sqrt{\langle\psi_0|e^{iH^\dagger t}e^{-iHt}|\psi_0\rangle}},
\label{Eq:TE}
\end{equation}
where the denominator ensures that the state remains normalised during the non-unitary time evolution. Alternate methods of using just the right eigenstates for the time evolution and observables have been also put forward recently \cite{Longhi:2023, QianTian:2024}. To avoid the computational cost of evaluating the exponential of large matrices directly, we decompose the initial state in terms of the right eigenstates $|V_n^R\rangle$ of the non-Hermitian Hamiltonian $\hat H$, which
satisfy
\begin{equation}
\hat H|V_n^R\rangle=E_n|V_n^R\rangle.
\end{equation}
Unlike the eigenstates of a Hermitian Hamiltonian, the right eigenstates are generally not orthogonal. We therefore introduce the left eigenstates $|V_n^L\rangle$, obtained from the Hermitian-conjugated Hamiltonian \cite{Hamazaki:2019},
\begin{equation}
\hat H^\dagger|V_n^L\rangle=E_n^*|V_n^L\rangle,
\end{equation}
which form a biorthogonal basis with the right eigenstates,
\begin{equation}
\langle V_n^L|V_m^R\rangle =
\delta_{nm}\langle V_n^L|V_n^R\rangle.
\end{equation}
In general, the overlap $\langle V_n^L|V_n^R\rangle \neq 1$ when the left and right eigenstates are normalised independently. In the numerical calculation, the right and left eigenstates are obtained by diagonalising $\hat H$ and $\hat H^\dagger$, respectively, and are paired according to the complex conjugate eigenvalues $E_n$ and $E_n^*$. Using this biorthogonal basis, the Hamiltonian can be decomposed as
\begin{equation}
\hat H=\sum_{n=1}^D E_n
\frac{|V_n^R\rangle\langle V_n^L|}
{\langle V_n^L|V_n^R\rangle},
\end{equation}
where $D$ is the dimension of the Fock basis. The numerator now of
Eq.~\eqref{Eq:TE} can then be written as
\begin{equation}
e^{-i\hat Ht}|\psi_0\rangle =
\sum_{n=1}^D e^{-iE_nt}
\frac{\langle V_n^L|\psi_0\rangle}
{\langle V_n^L|V_n^R\rangle}
|V_n^R\rangle.
\end{equation}
This spectral decomposition makes the evaluation of the non-Hermitian time evolution more efficient.

\section{Bloch oscillations under the Stark potential} 
\label{appendix:bloch_osc}
The one-dimensional hardcore boson bath can be mapped to non-interacting fermions through the Jordan--Wigner transformation. The corresponding single-particle Hamiltonian is
\begin{equation}
h_a=-J\sum_{n=1}^{L-1}\bigl[e^{-g}|n\rangle\langle n+1|+e^{+g}|n+1\rangle\langle n|\bigr]+\dA\sum_{n=1}^L n|n\rangle\langle n|.
\end{equation}
The $N_a$ particle many-body eigenstates are therefore Slater
determinants constructed from the single-particle eigenstates of
$h_a$. This allows the many-body bath dynamics to be related directly
to the corresponding single-particle problem. The matrix elements are $\langle n|h_a|n+1\rangle = -Je^{-g}$ and $\langle n+1|h_a|n\rangle = -Je^{+g}$, the Hatano--Nelson model. The right eigenvalue equation in position representation,
\begin{equation}\label{eq:eig_raw}
  \langle n|h_a|\psi^R\rangle = E\,\psi^R(n),
\end{equation}
expands to
\begin{equation}
-J e^{-g}\psi^R(n+1)-J e^{g}\psi^R(n-1)+\dA n\psi^R(n)=E\,\psi^R(n).
\end{equation}
Under open boundary conditions, we introduce the similarity transformation $\psi^R(n)=e^{gn}\varphi(n)$ gives
\begin{multline}
  -Je^{-g}e^{g(n+1)}\varphi(n+1) - Je^{g}e^{g(n-1)}\varphi(n-1) \\
  +\dA n\, e^{gn}\varphi(n) = E\,e^{gn}\varphi(n),
\end{multline}
\begin{multline}
  -Je^{gn}\varphi(n+1) - Je^{gn}\varphi(n-1) \\
  +\dA n\, e^{gn}\varphi(n) = E\,e^{gn}\varphi(n).
\end{multline}

\noindent The factors $e^{\pm g}$ from the asymmetric hopping exactly cancel, leaving
\begin{equation}\label{eq:wse}
  -J\biggl(\varphi(n+1)+\varphi(n-1)\biggr) +\dA n\, \varphi(n) = E\,\varphi(n).
\end{equation}
Equation \eqref{eq:wse} is the Hermitian Wannier--Stark equation with no dependence on $g$. The same similarity transformation extends directly to the full impurity-bath Hamiltonian. Defining
\begin{equation}
\hat S=\exp\left(g\sum_{i=1}^L i\,\hat n_i^a\right)\;,
\end{equation}
the bath Hamiltonian satisfies
\begin{equation}
\hat H_a(g)=\hat S\hat H_a(0)\hat S^{-1}.
\end{equation}
Since $\hat H_b$ acts only on the impurity and the density-density
interaction $\hat H_{ab}$ depends only on particle densities, both
commute with $\hat S$. Hence,
\begin{equation}
\hat H(g)=\hat S\hat H(0)\hat S^{-1}.
\end{equation}
Therefore, under OBC, the full many-body spectrum is independent of
$g$ and remains real for any $g$. The non-reciprocity consequently
changes the eigenstates and their dynamics without changing the
many-body eigenvalues.

On an infinite chain, its eigenstates are 
\begin{equation}
\varphi_\nu(n)=\mathcal{J}_{n-\nu}(2J/\dA),
\qquad
E_\nu=\dA \nu,
\end{equation}
giving the right and left eigenstates quoted in Eq.~\eqref{eq:eigenstates}. The spectrum is an equally spaced Wannier--Stark ladder \cite{Wannier:1962, Fukuyama:1973, Hart:1987}. The two localisation lengths in the eigenstates arise from different mechanisms. The Bessel-function envelope confines each Wannier--Stark state around its ladder site $\nu$, with a characteristic Stark localisation length $\ell_S=2J/\dA$. In contrast, the non-Hermitian factor $e^{\pm gn}$ biases the state towards one edge of the system, giving rise to the skin length $\ell_{\rm skin}=1/(2g)$. Since the gauge transformation depends only on $n$ and is independent of $\nu$, it acts as a similarity transformation on the full open-chain Hamiltonian. As a result, the eigenvalues are completely independent of $g$ for any system size $L$. The non-Hermitian hopping therefore changes the spatial profile of the eigenstates, but leaves the single-particle spectrum unchanged.

\begin{figure}
    \centering
    \includegraphics[width=\linewidth]{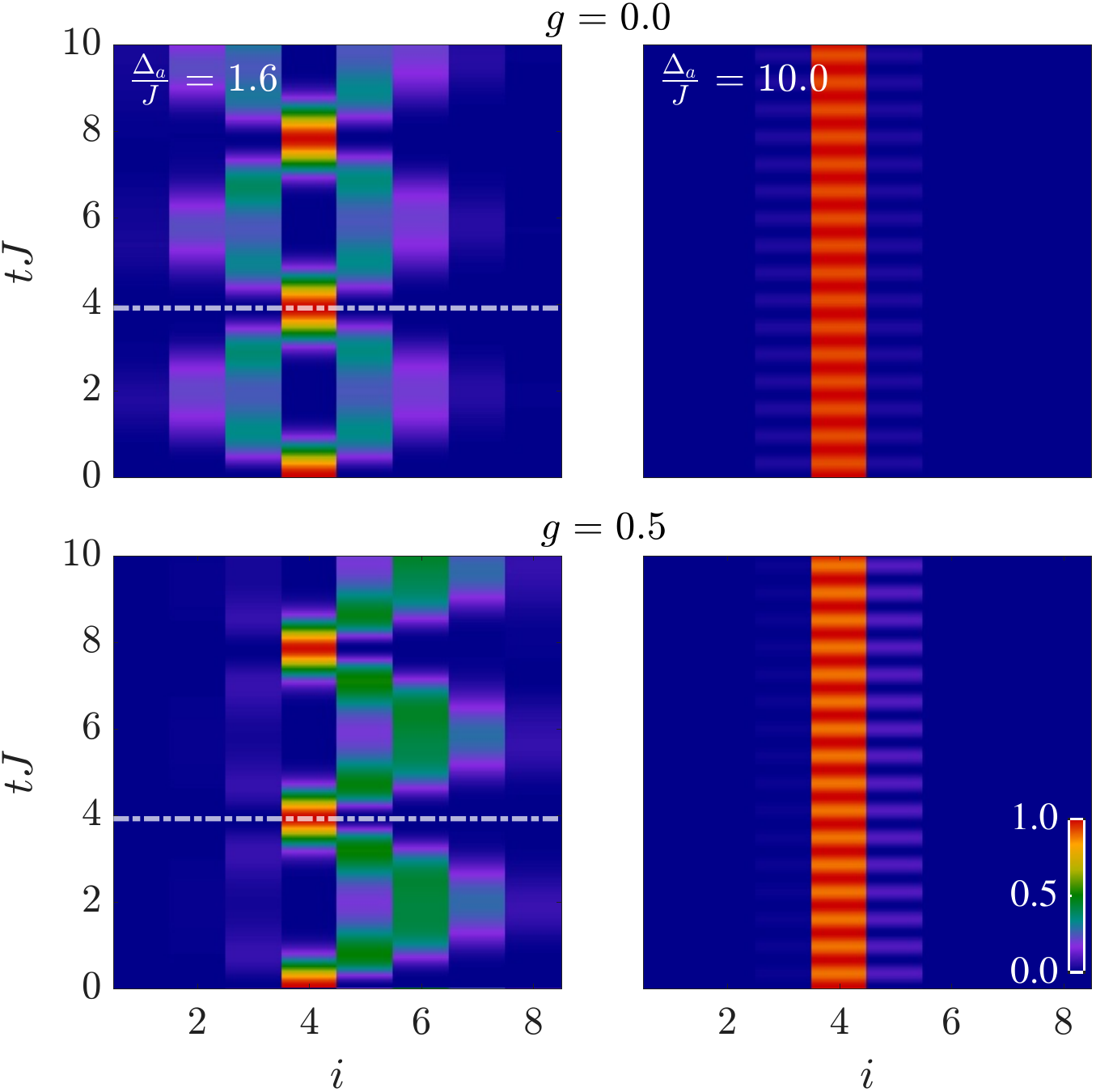}
    \caption{Single-particle dynamics in the Stark potential for a chain of $L=8$ sites, initially prepared at the central site. Top panels show Hermitian case $g=0$. For the smaller tilt $\dA=1.6J$ (top left), the particle undergoes Bloch oscillations with period $\tau_B=2\pi/\dA\simeq3.93$ (dashed line). For the larger tilt $\dA=10J$ (top right), the Stark localisation length $\ell_S=2J/\dA$ becomes much smaller than one lattice spacing and the particle density remains nearly stationary. Non-Hermitian case $g=0.5$ is shown in the bottom panels. At $\dA=1.6J$ (bottom left), the hopping asymmetry biases the oscillatory motion towards the system boundary through the non-Hermitian skin effect. At $\dA=10J$ (bottom right), the strong Stark localisation suppresses the dynamics and the density remains effectively frozen despite the asymmetric hopping as $g\ll\dA/J$.}
    \label{fig:bloch_osc}
\end{figure}

The Wannier--Stark ladder also determines the real-time dynamics of a single particle. Since the eigenvalues are equally spaced by $\dA$, a particle initially localised on a lattice site evolves and exhibits Bloch oscillations, a well-known phenomenon which has been studied extensively in various systems \cite{Bloch:1929, Dias:2007, Khomeriki:2010, Morsch:2016, Hasan:2022, Rabec:2025}. The particle oscillates with a period $\tau_B=\frac{2\pi}{\dA}$. The tilt therefore sets both the spatial extent of the Wannier--Stark orbitals, $\ell_S=2J/\dA$, and the timescale of their dynamics.

This behaviour is shown in top left panel of Fig. ~\ref{fig:bloch_osc}. For a relatively small tilt, $\dA=1.6J$, the single-particle density oscillates periodically across the lattice, with a period consistent with $\tau_B=2\pi/\dA \approx 3.927 $ shown as dashed line. The particle therefore remains dynamically localised within a finite region while undergoing repeated Bloch oscillations. As the tilt is increased to $\dA=10J$, the Stark localisation length becomes much smaller than the lattice spacing, $\ell_S=2J/\dA\ll1$, and tunnelling between neighbouring sites is strongly suppressed. The particle density then remains essentially stationary over time as shown in top right panel in Fig. \ref{fig:bloch_osc}.

The non-Hermitian asymmetry does not change the time period $\tau_B$ of the oscillations, since it leaves the Wannier--Stark spectrum unchanged. Instead, it modifies the spatial profile of the evolving state through the skin factor $e^{\pm g}$. This is clearly seen in bottom left panel in Fig.~\ref{fig:bloch_osc} where $g=0.5$ forces the particle to go towards the boundary. On the other hand, when $\dA=10\gg J$, the asymmetric hopping has little effect on the density which remains frozen in time. The single-particle dynamics therefore provides a direct illustration of the two competing length scales. the Stark length $\ell_S$ controls the spatial extent of the oscillatory motion, while the skin length $\ell_{\rm skin}$ controls its spatial bias towards the boundary.

% \section*{References}
\bibliography{references}

@article{Hart:1987,
  title = {Existence of {Wannier}-{Stark} localization},
  author = {Emin, David and Hart, C. F.},
  journal = {Phys. Rev. B},
  volume = {36},
  issue = {14},
  pages = {7353--7359},
  numpages = {0},
  year = {1987},
  month = {Nov},
  publisher = {American Physical Society},
  doi = {10.1103/PhysRevB.36.7353},
  url = {https://link.aps.org/doi/10.1103/PhysRevB.36.7353}
}

@article{Nandkishore:2015,
   author = "Nandkishore, Rahul and Huse, David A.",
   title = "Many-Body Localization and Thermalization in Quantum Statistical Mechanics", 
   journal= "Annual Review of Condensed Matter Physics",
   year = "2015",
   volume = "6",
   number = "Volume 6, 2015",
   pages = "15-38",
   doi = "https://doi.org/10.1146/annurev-conmatphys-031214-014726",
   url = "https://www.annualreviews.org/content/journals/10.1146/annurev-conmatphys-031214-014726",
   publisher = "Annual Reviews",
   issn = "1947-5462",
   type = "Journal Article",
  }

@article{Abanin:2019,
  title = {Colloquium: Many-body localization, thermalization, and entanglement},
  author = {Abanin, Dmitry A. and Altman, Ehud and Bloch, Immanuel and Serbyn, Maksym},
  journal = {Rev. Mod. Phys.},
  volume = {91},
  issue = {2},
  pages = {021001},
  numpages = {26},
  year = {2019},
  month = {May},
  publisher = {American Physical Society},
  doi = {10.1103/RevModPhys.91.021001},
  url = {https://link.aps.org/doi/10.1103/RevModPhys.91.021001}
}

@article{Schulz:2019,
  title = {Stark Many-Body Localization},
  author = {Schulz, M. and Hooley, C. A. and Moessner, R. and Pollmann, F.},
  journal = {Phys. Rev. Lett.},
  volume = {122},
  issue = {4},
  pages = {040606},
  numpages = {5},
  year = {2019},
  month = {Jan},
  publisher = {American Physical Society},
  doi = {10.1103/PhysRevLett.122.040606},
  url = {https://link.aps.org/doi/10.1103/PhysRevLett.122.040606}
}

@article{Nieuwenburg:2019,
   title={From {Bloch} oscillations to many-body localization in clean interacting systems},
   author={van Nieuwenburg, Evert and Baum, Yuval and Refael, Gil},
   journal={Proceedings of the National Academy of Sciences},
   volume={116},
   ISSN={1091-6490},
   number={19},
   publisher={National Academy of Sciences},
   year={2019},
    DOI={10.1073/pnas.1819316116},
   }

@article{Taylor:2020,
  title = {Experimental probes of {Stark} many-body localization},
  author = {Taylor, S. R. and Schulz, M. and Pollmann, F. and Moessner, R.},
  journal = {Phys. Rev. B},
  volume = {102},
  issue = {5},
  pages = {054206},
  numpages = {10},
  year = {2020},
  month = {Aug},
  publisher = {American Physical Society},
  doi = {10.1103/PhysRevB.102.054206},
  url = {https://link.aps.org/doi/10.1103/PhysRevB.102.054206}
}

@article{Guo:2021,
  title = {Stark Many-Body Localization on a Superconducting Quantum Processor},
  author = {Guo, Qiujiang and Cheng, Chen and Li, Hekang and Xu, Shibo and Zhang, Pengfei and Wang, Zhen and Song, Chao and Liu, Wuxin and Ren, Wenhui and Dong, Hang and Mondaini, Rubem and Wang, H.},
  journal = {Phys. Rev. Lett.},
  volume = {127},
  issue = {24},
  pages = {240502},
  numpages = {7},
  year = {2021},
  month = {Dec},
  publisher = {American Physical Society},
  doi = {10.1103/PhysRevLett.127.240502},
  url = {https://link.aps.org/doi/10.1103/PhysRevLett.127.240502}
}

@article{Brighi:2022,
  title = {Localization of a mobile impurity interacting with an Anderson insulator},
  author = {Brighi, Pietro and Michailidis, Alexios A. and Kirova, Kristina and Abanin, Dmitry A. and Serbyn, Maksym},
  journal = {Phys. Rev. B},
  volume = {105},
  issue = {22},
  pages = {224208},
  numpages = {18},
  year = {2022},
  month = {Jun},
  publisher = {American Physical Society},
  doi = {10.1103/PhysRevB.105.224208},
  url = {https://link.aps.org/doi/10.1103/PhysRevB.105.224208}
}

@article{Brighi:2022b,
  title = {Propagation of many-body localization in an {Anderson} insulator},
  author = {Brighi, Pietro and Michailidis, Alexios A. and Abanin, Dmitry A. and Serbyn, Maksym},
  journal = {Phys. Rev. B},
  volume = {105},
  issue = {22},
  pages = {L220203},
  numpages = {6},
  year = {2022},
  month = {Jun},
  publisher = {American Physical Society},
  doi = {10.1103/PhysRevB.105.L220203},
  url = {https://link.aps.org/doi/10.1103/PhysRevB.105.L220203}
}

@article{Brighi:2023,
  title = {Many-body localization proximity effect in a two-species bosonic Hubbard model},
  author = {Brighi, Pietro and Ljubotina, Marko and Abanin, Dmitry A. and Serbyn, Maksym},
  journal = {Phys. Rev. B},
  volume = {108},
  issue = {5},
  pages = {054201},
  numpages = {18},
  year = {2023},
  month = {Aug},
  publisher = {American Physical Society},
  doi = {10.1103/PhysRevB.108.054201},
  url = {https://link.aps.org/doi/10.1103/PhysRevB.108.054201}
}

@article{Falcao:2023,
  title = {Nonergodic dynamics for an impurity interacting with bosons in a tilted lattice},
  author = {Falc\~ao, Pedro R. Nic\'acio and Zakrzewski, Jakub},
  journal = {Phys. Rev. B},
  volume = {108},
  issue = {13},
  pages = {134201},
  numpages = {12},
  year = {2023},
  month = {Oct},
  publisher = {American Physical Society},
  doi = {10.1103/PhysRevB.108.134201},
  url = {https://link.aps.org/doi/10.1103/PhysRevB.108.134201}
}

@article{HatanoNelson:1996,
  title = {Localization Transitions in Non-Hermitian Quantum Mechanics},
  author = {Hatano, Naomichi and Nelson, David R.},
  journal = {Phys. Rev. Lett.},
  volume = {77},
  issue = {3},
  pages = {570--573},
  numpages = {0},
  year = {1996},
  month = {Jul},
  publisher = {American Physical Society},
  doi = {10.1103/PhysRevLett.77.570},
  url = {https://link.aps.org/doi/10.1103/PhysRevLett.77.570}
}

@article{HatanoNelson:1997,
  title = {Vortex pinning and non-Hermitian quantum mechanics},
  author = {Hatano, Naomichi and Nelson, David R.},
  journal = {Phys. Rev. B},
  volume = {56},
  issue = {14},
  pages = {8651--8673},
  numpages = {0},
  year = {1997},
  month = {Oct},
  publisher = {American Physical Society},
  doi = {10.1103/PhysRevB.56.8651},
  url = {https://link.aps.org/doi/10.1103/PhysRevB.56.8651}
}

@article{Ashida:2020,
   title={Non-Hermitian physics},
   volume={69},
   ISSN={1460-6976},
   url={http://dx.doi.org/10.1080/00018732.2021.1876991},
   DOI={10.1080/00018732.2021.1876991},
   number={3},
   journal={Advances in Physics},
   publisher={Informa UK Limited},
   author={Ashida, Yuto and Gong, Zongping and Ueda, Masahito},
   year={2020},
   month={Jul}, pages={249–435} }

@article{Lee:2016,
  title = {Anomalous Edge State in a Non-Hermitian Lattice},
  author = {Lee, Tony E.},
  journal = {Phys. Rev. Lett.},
  volume = {116},
  issue = {13},
  pages = {133903},
  numpages = {5},
  year = {2016},
  month = {Apr},
  publisher = {American Physical Society},
  doi = {10.1103/PhysRevLett.116.133903},
  url = {https://link.aps.org/doi/10.1103/PhysRevLett.116.133903}
}

@article{Liu:2023,
  title = {From ergodicity to many-body localization in a one-dimensional interacting non-Hermitian {Stark} system},
  author = {Liu, Jinghu and Xu, Zhihao},
  journal = {Phys. Rev. B},
  volume = {108},
  issue = {18},
  pages = {184205},
  numpages = {11},
  year = {2023},
  month = {Nov},
  publisher = {American Physical Society},
  doi = {10.1103/PhysRevB.108.184205},
  url = {https://link.aps.org/doi/10.1103/PhysRevB.108.184205}
}

@article{Lukin:2019,
   title={Probing entanglement in a many-body–localized system},
   volume={364},
   ISSN={1095-9203},
   url={http://dx.doi.org/10.1126/science.aau0818},
   DOI={10.1126/science.aau0818},
   number={6437},
   journal={Science},
   publisher={American Association for the Advancement of Science (AAAS)},
   author={Lukin, Alexander and Rispoli, Matthew and Schittko, Robert and Tai, M. Eric and Kaufman, Adam M. and Choi, Soonwon and Khemani, Vedika and Léonard, Julian and Greiner, Markus},
   year={2019},
   month=Apr, pages={256–260} }

@article{Xie:2024,
  title = {Determining the non-Hermitian parent Hamiltonian from a single eigenstate},
  author = {Xie, Xu-Dan and Xue, Zheng-Yuan and Zhang, Dan-Bo},
  journal = {Phys. Rev. B},
  volume = {110},
  issue = {23},
  pages = {235113},
  numpages = {8},
  year = {2024},
  month = {Dec},
  publisher = {American Physical Society},
  doi = {10.1103/PhysRevB.110.235113},
  url = {https://link.aps.org/doi/10.1103/PhysRevB.110.235113}
}

@article{Hatano:1998,
  title = {Non-Hermitian delocalization and eigenfunctions},
  author = {Hatano, Naomichi and Nelson, David R.},
  journal = {Phys. Rev. B},
  volume = {58},
  issue = {13},
  pages = {8384--8390},
  numpages = {0},
  year = {1998},
  month = {Oct},
  publisher = {American Physical Society},
  doi = {10.1103/PhysRevB.58.8384},
  url = {https://link.aps.org/doi/10.1103/PhysRevB.58.8384}
}

@article{Hatano:2021,
   title={Delocalization of a non-Hermitian quantum walk on random media in one dimension},
   volume={435},
   ISSN={0003-4916},
   url={http://dx.doi.org/10.1016/j.aop.2021.168615},
   DOI={10.1016/j.aop.2021.168615},
   journal={Annals of Physics},
   publisher={Elsevier BV},
   author={Hatano, Naomichi and Obuse, Hideaki},
   year={2021},
   month=Dec, pages={168615} 
   }

@article{Zeng:2023,
   title={Wannier-{Stark} localization in one-dimensional amplitude-chirped lattices},
   volume={108},
   ISSN={2469-9969},
   url={http://dx.doi.org/10.1103/PhysRevB.108.104207},
   DOI={10.1103/physrevb.108.104207},
   number={10},
   journal={Physical Review B},
   publisher={American Physical Society (APS)},
   author={Zeng, Qi-Bo and Hou, Bo and Xiao, Han},
   year={2023},
   month={Sept} }

@article{Gornyi:2005,
  title = {Interacting Electrons in Disordered Wires: {Anderson} Localization and Low-$T$ Transport},
  author = {Gornyi, I. V. and Mirlin, A. D. and Polyakov, D. G.},
  journal = {Phys. Rev. Lett.},
  volume = {95},
  issue = {20},
  pages = {206603},
  numpages = {4},
  year = {2005},
  month = {Nov},
  publisher = {American Physical Society},
  doi = {10.1103/PhysRevLett.95.206603},
  url = {https://link.aps.org/doi/10.1103/PhysRevLett.95.206603}
}

@ARTICLE{Anderson:1958,
       author = {{Anderson}, P.~W.},
        title = "{Absence of Diffusion in Certain Random Lattices}",
      journal = {Physical Review},
         year = 1958,
        month = mar,
       volume = {109},
       number = {5},
        pages = {1492-1505},
          doi = {10.1103/PhysRev.109.1492}
}

@article{Basko:2006,
title = {Metal–insulator transition in a weakly interacting many-electron system with localized single-particle states},
journal = {Annals of Physics},
volume = {321},
number = {5},
pages = {1126-1205},
year = {2006},
issn = {0003-4916},
doi = {https://doi.org/10.1016/j.aop.2005.11.014},
url = {https://www.sciencedirect.com/science/article/pii/S0003491605002630},
author = {D.M. Basko and I.L. Aleiner and B.L. Altshuler}
}

@article{Serbyn:2013,
  title = {Local Conservation Laws and the Structure of the Many-Body Localized States},
  author = {Serbyn, Maksym and Papi\ifmmode \acute{c}\else \'{c}\fi{}, Z. and Abanin, Dmitry A.},
  journal = {Phys. Rev. Lett.},
  volume = {111},
  issue = {12},
  pages = {127201},
  numpages = {5},
  year = {2013},
  month = {Sep},
  publisher = {American Physical Society},
  doi = {10.1103/PhysRevLett.111.127201},
  url = {https://link.aps.org/doi/10.1103/PhysRevLett.111.127201}
}

@article{Huse:2014,
  title = {Phenomenology of fully many-body-localized systems},
  author = {Huse, David A. and Nandkishore, Rahul and Oganesyan, Vadim},
  journal = {Phys. Rev. B},
  volume = {90},
  issue = {17},
  pages = {174202},
  numpages = {5},
  year = {2014},
  month = {Nov},
  publisher = {American Physical Society},
  doi = {10.1103/PhysRevB.90.174202},
  url = {https://link.aps.org/doi/10.1103/PhysRevB.90.174202}
}

@article{Imbrie:2016,
   title={On Many-Body Localization for Quantum Spin Chains},
   volume={163},
   ISSN={1572-9613},
   url={http://dx.doi.org/10.1007/s10955-016-1508-x},
   DOI={10.1007/s10955-016-1508-x},
   number={5},
   journal={Journal of Statistical Physics},
   publisher={Springer Science and Business Media LLC},
   author={Imbrie, John Z.},
   year={2016},
   month=Apr, pages={998–1048} }

@article{Schreiber:2015,
author = {Michael Schreiber  and Sean S. Hodgman  and Pranjal Bordia  and Henrik P. Lüschen  and Mark H. Fischer  and Ronen Vosk  and Ehud Altman  and Ulrich Schneider  and Immanuel Bloch },
title = {Observation of many-body localization of interacting fermions in a quasirandom optical lattice},
journal = {Science},
volume = {349},
number = {6250},
pages = {842-845},
year = {2015},
doi = {10.1126/science.aaa7432},
URL = {https://www.science.org/doi/abs/10.1126/science.aaa7432},
}

@article{Luschen:2017,
  title = {Observation of Slow Dynamics near the Many-Body Localization Transition in One-Dimensional Quasiperiodic Systems},
  author = {L\"uschen, Henrik P. and Bordia, Pranjal and Scherg, Sebastian and Alet, Fabien and Altman, Ehud and Schneider, Ulrich and Bloch, Immanuel},
  journal = {Phys. Rev. Lett.},
  volume = {119},
  issue = {26},
  pages = {260401},
  numpages = {6},
  year = {2017},
  month = {Dec},
  publisher = {American Physical Society},
  doi = {10.1103/PhysRevLett.119.260401},
  url = {https://link.aps.org/doi/10.1103/PhysRevLett.119.260401}
}

@article{Weiner:2019,
  title = {Slow dynamics and strong finite-size effects in many-body localization with random and quasiperiodic potentials},
  author = {Weiner, Felix and Evers, Ferdinand and Bera, Soumya},
  journal = {Phys. Rev. B},
  volume = {100},
  issue = {10},
  pages = {104204},
  numpages = {11},
  year = {2019},
  month = {Sep},
  publisher = {American Physical Society},
  doi = {10.1103/PhysRevB.100.104204},
  url = {https://link.aps.org/doi/10.1103/PhysRevB.100.104204}
}

@article{Bardarson:2012,
  title = {Unbounded Growth of Entanglement in Models of Many-Body Localization},
  author = {Bardarson, Jens H. and Pollmann, Frank and Moore, Joel E.},
  journal = {Phys. Rev. Lett.},
  volume = {109},
  issue = {1},
  pages = {017202},
  numpages = {5},
  year = {2012},
  month = {Jul},
  publisher = {American Physical Society},
  doi = {10.1103/PhysRevLett.109.017202},
  url = {https://link.aps.org/doi/10.1103/PhysRevLett.109.017202}
}

@article{Serbyn:2013b,
  title = {Universal Slow Growth of Entanglement in Interacting Strongly Disordered Systems},
  author = {Serbyn, Maksym and Papi\ifmmode \acute{c}\else \'{c}\fi{}, Z. and Abanin, Dmitry A.},
  journal = {Phys. Rev. Lett.},
  volume = {110},
  issue = {26},
  pages = {260601},
  numpages = {5},
  year = {2013},
  month = {Jun},
  publisher = {American Physical Society},
  doi = {10.1103/PhysRevLett.110.260601},
  url = {https://link.aps.org/doi/10.1103/PhysRevLett.110.260601}
}

@article{Bera:2017,
  title = {Density Propagator for Many-Body Localization: Finite-Size Effects, Transient Subdiffusion, and Exponential Decay},
  author = {Bera, Soumya and De Tomasi, Giuseppe and Weiner, Felix and Evers, Ferdinand},
  journal = {Phys. Rev. Lett.},
  volume = {118},
  issue = {19},
  pages = {196801},
  numpages = {6},
  year = {2017},
  month = {May},
  publisher = {American Physical Society},
  doi = {10.1103/PhysRevLett.118.196801},
  url = {https://link.aps.org/doi/10.1103/PhysRevLett.118.196801}
}

@article{Guo:2020,
   title={Observation of energy-resolved many-body localization},
   volume={17},
   ISSN={1745-2481},
   url={http://dx.doi.org/10.1038/s41567-020-1035-1},
   DOI={10.1038/s41567-020-1035-1},
   number={2},
   journal={Nature Physics},
   publisher={Springer Science and Business Media LLC},
   author={Guo, Qiujiang and Cheng, Chen and Sun, Zheng-Hang and Song, Zixuan and Li, Hekang and Wang, Zhen and Ren, Wenhui and Dong, Hang and Zheng, Dongning and Zhang, Yu-Ran and Mondaini, Rubem and Fan, Heng and Wang, H.},
   year={2020},
   month={Sept}, pages={234–239} }

@article{Nandkishore:2015b,
  title = {Many-body localization proximity effect},
  author = {Nandkishore, Rahul},
  journal = {Phys. Rev. B},
  volume = {92},
  issue = {24},
  pages = {245141},
  numpages = {5},
  year = {2015},
  month = {Dec},
  publisher = {American Physical Society},
  doi = {10.1103/PhysRevB.92.245141}
}

@article{Luitz:2017,
  title = {How a Small Quantum Bath Can Thermalize Long Localized Chains},
  author = {Luitz, David J. and Huveneers, Fran\ifmmode \mbox{\c{c}}\else \c{c}\fi{}ois and De Roeck, Wojciech},
  journal = {Phys. Rev. Lett.},
  volume = {119},
  issue = {15},
  pages = {150602},
  numpages = {6},
  year = {2017},
  month = {Oct},
  publisher = {American Physical Society},
  doi = {10.1103/PhysRevLett.119.150602},
  url = {https://link.aps.org/doi/10.1103/PhysRevLett.119.150602}
}

@article{Wybo:2020,
  title = {Entanglement dynamics of a many-body localized system coupled to a bath},
  author = {Wybo, Elisabeth and Knap, Michael and Pollmann, Frank},
  journal = {Phys. Rev. B},
  volume = {102},
  issue = {6},
  pages = {064304},
  numpages = {11},
  year = {2020},
  month = {Aug},
  publisher = {American Physical Society},
  doi = {10.1103/PhysRevB.102.064304},
  url = {https://link.aps.org/doi/10.1103/PhysRevB.102.064304}
}

@article{Bender:1998,
  title = {Real Spectra in Non-Hermitian Hamiltonians Having $\mathcal{P}\mathcal{T}$ Symmetry},
  author = {Bender, Carl M. and Boettcher, Stefan},
  journal = {Phys. Rev. Lett.},
  volume = {80},
  issue = {24},
  pages = {5243--5246},
  numpages = {0},
  year = {1998},
  month = {Jun},
  publisher = {American Physical Society},
  doi = {10.1103/PhysRevLett.80.5243},
  url = {https://link.aps.org/doi/10.1103/PhysRevLett.80.5243}
}

@article{Longhi:2009,
  title = {Bloch Oscillations in Complex Crystals with $\mathcal{P}\mathcal{T}$ Symmetry},
  author = {Longhi, S.},
  journal = {Phys. Rev. Lett.},
  volume = {103},
  issue = {12},
  pages = {123601},
  numpages = {4},
  year = {2009},
  month = {Sep},
  publisher = {American Physical Society},
  doi = {10.1103/PhysRevLett.103.123601},
  url = {https://link.aps.org/doi/10.1103/PhysRevLett.103.123601}
}

@article{Ashida:2017,
   title={Parity-time-symmetric quantum critical phenomena},
   volume={8},
   ISSN={2041-1723},
   url={http://dx.doi.org/10.1038/ncomms15791},
   DOI={10.1038/ncomms15791},
   number={1},
   journal={Nature Communications},
   publisher={Springer Science and Business Media LLC},
   author={Ashida, Yuto and Furukawa, Shunsuke and Ueda, Masahito},
   year={2017},
   month={June} }

@article{Xiao:2020,
   title={Non-Hermitian bulk–boundary correspondence in quantum dynamics},
   volume={16},
   ISSN={1745-2481},
   url={http://dx.doi.org/10.1038/s41567-020-0836-6},
   DOI={10.1038/s41567-020-0836-6},
   number={7},
   journal={Nature Physics},
   publisher={Springer Science and Business Media LLC},
   author={Xiao, Lei and Deng, Tianshu and Wang, Kunkun and Zhu, Gaoyan and Wang, Zhong and Yi, Wei and Xue, Peng},
   year={2020},
   month=Mar, pages={761–766} }

@article{Ji:2024,
   title={Generalized bulk-boundary correspondence in periodically driven non-Hermitian systems},
   volume={36},
   ISSN={1361-648X},
   url={http://dx.doi.org/10.1088/1361-648X/ad2c73},
   DOI={10.1088/1361-648x/ad2c73},
   number={24},
   journal={Journal of Physics: Condensed Matter},
   publisher={IOP Publishing},
   author={Ji, Xiang and Yang, Xiaosen},
   year={2024},
   month=Mar, pages={243001} }

@article{Hofmann:2020,
  title = {Reciprocal skin effect and its realization in a topolectrical circuit},
  author = {Hofmann, Tobias and Helbig, Tobias and Schindler, Frank and Salgo, Nora and Brzezi\ifmmode \acute{n}\else \'{n}\fi{}ska, Marta and Greiter, Martin and Kiessling, Tobias and Wolf, David and Vollhardt, Achim and Kaba\ifmmode \check{s}\else \v{s}\fi{}i, Anton and Lee, Ching Hua and Bilu\ifmmode \check{s}\else \v{s}\fi{}i\ifmmode \acute{c}\else \'{c}\fi{}, Ante and Thomale, Ronny and Neupert, Titus},
  journal = {Phys. Rev. Res.},
  volume = {2},
  issue = {2},
  pages = {023265},
  numpages = {11},
  year = {2020},
  month = {Jun},
  publisher = {American Physical Society},
  doi = {10.1103/PhysRevResearch.2.023265},
  url = {https://link.aps.org/doi/10.1103/PhysRevResearch.2.023265}
}

@article{WeiGou:2020,
  title = {Tunable Nonreciprocal Quantum Transport through a Dissipative {Aharonov}-{Bohm} Ring in Ultracold Atoms},
  author = {Gou, Wei and Chen, Tao and Xie, Dizhou and Xiao, Teng and Deng, Tian-Shu and Gadway, Bryce and Yi, Wei and Yan, Bo},
  journal = {Phys. Rev. Lett.},
  volume = {124},
  issue = {7},
  pages = {070402},
  numpages = {6},
  year = {2020},
  month = {Feb},
  publisher = {American Physical Society},
  doi = {10.1103/PhysRevLett.124.070402},
  url = {https://link.aps.org/doi/10.1103/PhysRevLett.124.070402}
}

@article{Liang:2022,
  title = {Dynamic Signatures of Non-Hermitian Skin Effect and Topology in Ultracold Atoms},
  author = {Liang, Qian and Xie, Dizhou and Dong, Zhaoli and Li, Haowei and Li, Hang and Gadway, Bryce and Yi, Wei and Yan, Bo},
  journal = {Phys. Rev. Lett.},
  volume = {129},
  issue = {7},
  pages = {070401},
  numpages = {6},
  year = {2022},
  month = {Aug},
  publisher = {American Physical Society},
  doi = {10.1103/PhysRevLett.129.070401},
  url = {https://link.aps.org/doi/10.1103/PhysRevLett.129.070401}
}

@article{Li:2023,
   title={Non-Hermitian {Stark} many-body localization},
   volume={108},
   ISSN={2469-9934},
   url={http://dx.doi.org/10.1103/PhysRevA.108.043301},
   DOI={10.1103/physreva.108.043301},
   number={4},
   journal={Physical Review A},
   publisher={American Physical Society (APS)},
   author={Li, Han-Ze and Yu, Xue-Jia and Zhong, Jian-Xin},
   year={2023},
   month=Oct }

@article{Wang:2023,
   title={Non-Hermitian skin effects on thermal and many-body localized phases},
   volume={107},
   ISSN={2469-9969},
   url={http://dx.doi.org/10.1103/PhysRevB.107.L220205},
   DOI={10.1103/physrevb.107.l220205},
   number={22},
   journal={Physical Review B},
   publisher={American Physical Society (APS)},
   author={Wang, Yi-Cheng and Suthar, Kuldeep and Jen, H. H. and Hsu, Yi-Ting and You, Jhih-Shih},
   year={2023},
   month={June} }

@article{Chakrabarty:2026,
   title={Many-body non-Hermitian quasiperiodic systems with power-law hopping: Localization, topology, and the skin effect},
   volume={114},
   ISSN={2469-9969},
   url={http://dx.doi.org/10.1103/wl2m-8l4t},
   DOI={10.1103/wl2m-8l4t},
   number={3},
   journal={Physical Review B},
   publisher={American Physical Society (APS)},
   author={Chakrabarty, Aditi and Banerjee, Sanchayan and Mishra, Tapan and Datta, Sanjoy},
   year={2026},
   month={July} }

@article{Suthar:2022,
  title = {Non-Hermitian many-body localization with open boundaries},
  author = {Suthar, Kuldeep and Wang, Yi-Cheng and Huang, Yi-Ping and Jen, H. H. and You, Jhih-Shih},
  journal = {Phys. Rev. B},
  volume = {106},
  issue = {6},
  pages = {064208},
  numpages = {11},
  year = {2022},
  month = {Aug},
  publisher = {American Physical Society},
  doi = {10.1103/PhysRevB.106.064208},
  url = {https://link.aps.org/doi/10.1103/PhysRevB.106.064208}
}

@article{Luschen:2017b,
  title = {Signatures of Many-Body Localization in a Controlled Open Quantum System},
  author = {L\"uschen, Henrik P. and Bordia, Pranjal and Hodgman, Sean S. and Schreiber, Michael and Sarkar, Saubhik and Daley, Andrew J. and Fischer, Mark H. and Altman, Ehud and Bloch, Immanuel and Schneider, Ulrich},
  journal = {Phys. Rev. X},
  volume = {7},
  issue = {1},
  pages = {011034},
  numpages = {13},
  year = {2017},
  month = {Mar},
  publisher = {American Physical Society},
  doi = {10.1103/PhysRevX.7.011034},
  url = {https://link.aps.org/doi/10.1103/PhysRevX.7.011034}
}

@article{Morong:2021,
   title={Observation of {Stark} many-body localization without disorder},
   volume={599},
   ISSN={1476-4687},
   url={http://dx.doi.org/10.1038/s41586-021-03988-0},
   DOI={10.1038/s41586-021-03988-0},
   number={7885},
   journal={Nature},
   publisher={Springer Science and Business Media LLC},
   author={Morong, W. and Liu, F. and Becker, P. and Collins, K. S. and Feng, L. and Kyprianidis, A. and Pagano, G. and You, T. and Gorshkov, A. V. and Monroe, C.},
   year={2021},
   month=Nov, pages={393–398} }

@article{Shulin:2026,
  title = {Non-Hermitian Anomalous Scaling Engineering},
  author = {Wang, Shulin and He, Jiawei and Yang, Zhiyuan and Longhi, Stefano and Xue, Peng},
  journal = {Phys. Rev. Lett.},
  volume = {137},
  issue = {8},
  pages = {083801},
  numpages = {7},
  year = {2026},
  month = {Aug},
  publisher = {American Physical Society},
  doi = {10.1103/39t1-94yh},
  url = {https://link.aps.org/doi/10.1103/39t1-94yh}
}

@article{Hamazaki:2019,
  title = {Non-Hermitian Many-Body Localization},
  author = {Hamazaki, Ryusuke and Kawabata, Kohei and Ueda, Masahito},
  journal = {Phys. Rev. Lett.},
  volume = {123},
  issue = {9},
  pages = {090603},
  numpages = {7},
  year = {2019},
  month = {Aug},
  publisher = {American Physical Society},
  doi = {10.1103/PhysRevLett.123.090603}
}

@ARTICLE{Bloch:1929,
       author = {{Bloch}, Felix},
        title = "{{\"U}ber die Quantenmechanik der Elektronen in Kristallgittern}",
      journal = {Zeitschrift fur Physik},
         year = 1929,
        month = jul,
       volume = {52},
       number = {7-8},
        pages = {555-600},
          doi = {10.1007/BF01339455} }

@article{Dias:2007,
  title = {Frequency doubling of {Bloch} oscillations for interacting electrons in a static electric field},
  author = {Dias, W. S. and Nascimento, E. M. and Lyra, M. L. and de Moura, F. A. B. F.},
  journal = {Phys. Rev. B},
  volume = {76},
  issue = {15},
  pages = {155124},
  numpages = {6},
  year = {2007},
  month = {Oct},
  publisher = {American Physical Society},
  doi = {10.1103/PhysRevB.76.155124},
  url = {https://link.aps.org/doi/10.1103/PhysRevB.76.155124}
}

@article{Khomeriki:2010,
  title = {Interaction-induced fractional {Bloch} and tunneling oscillations},
  author = {Khomeriki, Ramaz and Krimer, Dmitry O. and Haque, Masudul and Flach, Sergej},
  journal = {Phys. Rev. A},
  volume = {81},
  issue = {6},
  pages = {065601},
  numpages = {4},
  year = {2010},
  month = {Jun},
  publisher = {American Physical Society},
  doi = {10.1103/PhysRevA.81.065601},
  url = {https://link.aps.org/doi/10.1103/PhysRevA.81.065601}
}

@article{Rabec:2025,
   title={Bloch oscillations of a soliton in a one-dimensional quantum fluid},
   volume={21},
   ISSN={1745-2481},
   url={http://dx.doi.org/10.1038/s41567-025-02970-1},
   DOI={10.1038/s41567-025-02970-1},
   number={10},
   journal={Nature Physics},
   publisher={Springer Science and Business Media LLC},
   author={Rabec, F. and Chauveau, G. and Brochier, G. and Nascimbene, S. and Dalibard, J. and Beugnon, J.},
   year={2025},
   month=aug, pages={1541–1547} 
}

@article{Hasan:2022,
doi = {10.1088/1361-6455/ac6ea3},
url = {https://doi.org/10.1088/1361-6455/ac6ea3},
year = {2022},
month = {may},
publisher = {IOP Publishing},
volume = {55},
number = {13},
pages = {135302},
author = {Hasan, Muhammad S and Polo, J and Pelayo, J C and Busch, Th},
title = {Bloch oscillations in supersolids},
journal = {Journal of Physics B: Atomic, Molecular and Optical Physics},
}

@article{Morsch:2016,
doi = {10.1088/1367-2630/18/11/111005},
year = {2016},
month = {11},
publisher = {IOP Publishing},
volume = {18},
number = {11},
pages = {111005},
author = {Morsch, Oliver},
title = {Taking a peek at {Bloch} oscillations},
journal = {New Journal of Physics},
}

@article{Daley:2014,
author = {Andrew J. Daley},
title = {Quantum trajectories and open many-body quantum systems},
journal = {Advances in Physics},
volume = {63},
number = {2},
pages = {77--149},
year = {2014},
publisher = {Taylor \& Francis},
doi = {10.1080/00018732.2014.933502}
}

@article{Brody:2014,
doi = {10.1088/1751-8113/47/3/035305},
url = {https://doi.org/10.1088/1751-8113/47/3/035305},
year = {2013},
month = {dec},
publisher = {IOP Publishing},
volume = {47},
number = {3},
pages = {035305},
author = {Brody, Dorje C},
title = {Biorthogonal quantum mechanics},
journal = {Journal of Physics A: Mathematical and Theoretical}
}

@article{Longhi:2023,
  title = {Phase transitions and bunching of correlated particles in a non-Hermitian quasicrystal},
  author = {Longhi, Stefano},
  journal = {Phys. Rev. B},
  volume = {108},
  issue = {7},
  pages = {075121},
  numpages = {12},
  year = {2023},
  month = {Aug},
  publisher = {American Physical Society},
  doi = {10.1103/PhysRevB.108.075121},
  url = {https://link.aps.org/doi/10.1103/PhysRevB.108.075121}
}

@article{QianTian:2024,
  title = {Correlation-induced phase transitions and mobility edges in an interacting non-Hermitian quasicrystal},
  author = {Qian, Tian and Gu, Yongjian and Zhou, Longwen},
  journal = {Phys. Rev. B},
  volume = {109},
  issue = {5},
  pages = {054204},
  numpages = {19},
  year = {2024},
  month = {Feb},
  publisher = {American Physical Society},
  doi = {10.1103/PhysRevB.109.054204},
  url = {https://link.aps.org/doi/10.1103/PhysRevB.109.054204}
}

@article{Wannier:1962,
  title = {Dynamics of Band Electrons in Electric and Magnetic Fields},
  author = {Wannier, Gregory H.},
  journal = {Rev. Mod. Phys.},
  volume = {34},
  issue = {4},
  pages = {645--655},
  numpages = {0},
  year = {1962},
  month = {Oct},
  publisher = {American Physical Society},
  doi = {10.1103/RevModPhys.34.645},
  url = {https://link.aps.org/doi/10.1103/RevModPhys.34.645}
}

@article{Fukuyama:1973,
  title = {Tightly Bound Electrons in a Uniform Electric Field},
  author = {Fukuyama, Hidetoshi and Bari, Robert A. and Fogedby, Hans C.},
  journal = {Phys. Rev. B},
  volume = {8},
  issue = {12},
  pages = {5579--5586},
  numpages = {0},
  year = {1973},
  month = {Dec},
  publisher = {American Physical Society},
  doi = {10.1103/PhysRevB.8.5579},
  url = {https://link.aps.org/doi/10.1103/PhysRevB.8.5579}
}

@article{YalunZhang:2025,
  title = {Two-body interaction induced phase transitions and intermediate phases in nonreciprocal non-Hermitian quasicrystals},
  author = {Zhang, Yalun and Zhou, Longwen},
  journal = {Phys. Rev. B},
  volume = {111},
  issue = {6},
  pages = {064204},
  numpages = {12},
  year = {2025},
  month = {Feb},
  publisher = {American Physical Society},
  doi = {10.1103/PhysRevB.111.064204},
  url = {https://link.aps.org/doi/10.1103/PhysRevB.111.064204}
}

@misc{QiRui:2023,
      title={Localization and mobility edges in non-Hermitian disorder-free lattices}, 
      author={Rui Qi and Junpeng Cao and Xiang-Ping Jiang},
      year={2023},
      eprint={2306.03807},
      archivePrefix={arXiv},
      primaryClass={cond-mat.dis-nn},
      url={https://arxiv.org/abs/2306.03807}, 
}
% \textit{References.}---

\end{document}